\documentclass[a4paper,11pt]{article}
\usepackage{titlesec}
\titleformat{\paragraph}[runin]{\normalfont\itshape}{\theparagraph.}{.3em}{}[.]\titlespacing{\paragraph}{0pt}{1ex plus .1ex minus .2ex}{.5em}
\usepackage{amsmath,amssymb,bm}
\usepackage{mathabx}
\usepackage[T1]{fontenc}
\usepackage[utf8]{inputenc}
\usepackage{lmodern}
\usepackage{mathtools}
\usepackage{dsfont}
\pdfoutput=1
\usepackage[english]{babel}
\usepackage[letterpaper, hmargin=1in, top=1in, bottom=1.2in, footskip=0.6in]{geometry}
\titleformat{\paragraph}[runin]{\normalfont\itshape}{\theparagraph.}{.3em}{}[.]\titlespacing{\paragraph}{0pt}{1ex plus .1ex minus .2ex}{.5em}
\usepackage[letterpaper, hmargin=1in, top=1in, bottom=1.2in, footskip=0.6in]{geometry}
\usepackage[utf8]{inputenc}
\usepackage{lmodern}
\usepackage{mathtools}
\usepackage{dsfont}
\pdfoutput=1
\usepackage[T1]{fontenc}
\usepackage[english]{babel}
\usepackage{graphicx} 
\usepackage{booktabs} 
\usepackage{color}
\definecolor{aquamarine}{rgb}{0.5, 1.0, 0.83}
\definecolor{ao(english)}{rgb}{0.0, 0.5, 0.0}
\definecolor{armygreen}{rgb}{0.29, 0.33, 0.13}
\definecolor{awesome}{rgb}{1.0, 0.13, 0.32}
\definecolor{ballblue}{rgb}{0.13, 0.67, 0.8}
\definecolor{bittersweet}{rgb}{1.0, 0.44, 0.37}
\definecolor{blue}{rgb}{0.0, 0.0, 1.0}
\definecolor{brinkpink}{rgb}{0.98, 0.38, 0.5}
\definecolor{ballblue}{rgb}{0.13, 0.67, 0.8}
\definecolor{brightturquoise}{rgb}{0.03, 0.91, 0.87}
\definecolor{blue-green}{rgb}{0.0, 0.87, 0.87}
\definecolor{caribbeangreen}{rgb}{0.0, 0.8, 0.6}
\definecolor{cyan}{rgb}{0.0, 1.0, 1.0}
\definecolor{amber(sae/ece)}{rgb}{1.0, 0.49, 0.0}
\graphicspath{ {images/} }
\usepackage[utf8]{inputenc}
\usepackage{lmodern}

\author{J\"{u}rg Fr\"{o}hlich$^{1}$ \quad Zhou Gang$^{2}$ \quad Alessandro Pizzo$^{3}$}

\title{A Completion of Quantum Mechanics\footnote{to appear in `Encyclopedia of Mathematical Physics'}}

\begin{document}

\maketitle

\begin{abstract}
A proposal of how to complete non-relativistic quantum mechanics to a physically 
meaningful, mathematically precise and logically coherent theory is reviewed. Our proposal 
leads to a general, non-linear stochastic law for the time-evolution of states of \textbf{individual}
physical systems. An application of the general formalism to the quantum theory of fluorescence 
of an atom coupled to the radiation field is sketched. Some remarks on relativistic quantum 
theory conclude our review.
\end{abstract}

\noindent Key words: Text-book quantum mechanics, Copenhagen interpretation, completion of quantum mechanics,
$ETH$ - Approach, potential events, actual events, Principle of Diminishing Potentialities, Huygens' principle,
state-reduction postulate, arrow of time, fluorescence\\

\noindent Key points:
\begin{itemize}
\item{The shortcomings of text-book quantum mechanics (QM) are discussed, and it is argued that they 
cannot be eliminated by an improved ``interpretation,'' but that a completion of the theory is necessary.}
 \item{It is emphasized that, in QM, the evolution of individual physical systems is non-linear and
stochastic; whereas the evolution of averages of states over a large ensemble of identical systems
is linear and deterministic, as claimed in standard  text-books.}
\item{In support of this claim, an analogy between QM, on one hand, and the theory of diffusion and Brownian motion,
on the other hand, is sketched.}
\item{The notions of potential events (future potentialities), events actualizing and actual events (actualities) 
are introduced into QM, and a general principle, called Principle of Diminishing Potentialities (PDP), is proposed, 
which states that the set of future potentialities shrinks with time.}
\item{A fundamental law determining the stochastic evolution of states of \textbf{individual} systems is derived
from PDP and a state-reduction postulate. These postulates represent the backbone of what has become 
known as the $ETH$ - Approach to QM. They supply a plausible ``ontology'' to QM.}
\item{It is explained how PDP can be derived from a quantum-mechanical version of Huygens' principle.}
\item{As an example of an application of the $ETH$ - Approach the theory of fluorescence is sketched.}
\item{Some elements of relativistic quantum theory are outlined, and it is pointed out that a completion of
QM that solves problems such as the ``measurement problem'' or the ``information paradox'' appears to 
be necessarily a local quantum theory with infinitely many degrees of freedom including massless ones.}
\end{itemize}

\section{What is missing in conventional quantum mechanics (QM)?}\label{Intro}

The frame of mind underlying this review can be succinctly characterized by quoting Paul Adrien Maurice 
\textit{Dirac,} who said: \textit{``The interpretation of quantum mechanics has been dealt with by many authors, 
and I do not want to discuss it here. I want to deal with more fundamental things;''} and: \textit{``It seems 
clear that the present quantum mechanics is not in its final form.''} 

This paper does not contain yet another interpretation of QM; it contains an outline
of a proposal of how to extend text-book QM to a theory that is logically coherent and gives 
an accurate account of processes in Nature; i.e., a \textbf{``quantum mechanics in its final form.''} 
Among the main features of our proposal there are the following ones.
\begin{enumerate}
\item{The average of states of physical systems taken over a large ensemble of \textbf{identical, 
identically prepared systems}, which we call an \textbf{``ensemble state''}, evolves in time according 
to a linear, deterministic evolution equation -- for isolated systems, a Schr\"odinger - von Neumann equation.
This is in accordance with text-book QM.}
\item{In contrast, the evolution of the state of an \textbf{individual} system is usually (non-linear and) 
\textbf{stochastic}; it is \textbf{not} described by a Schr\"odinger equation. Our task is to find the fundamental 
quantum-mechanical \textbf{law} that determines the stochastic time evolution of individual physical systems. 
It turns out that this law incorporates a fundamental ``arrow of time.'' In order to formulate it, it will turn out to
be  necessary to first clarify what ``potential events'' and ``actual events'' are in QM and how they express 
the dichotomy between past and future.}
\item{\textbf{Conservation laws}, in particular energy conservation, only hold for large ensembles of identical
systems, i.e., for the time evolution of \textbf{ensemble states}; but they do \textbf{not} hold strictly for 
individual systems. This will be seen to be a consequence of the stochastic evolution of the states of
individual systems, and in particular of the state-reduction postulate formulated in Section 3.}
\item{The basic reason why individual systems exhibit stochastic evolution appears to be that they 
contain matter degrees of freedom coupled to fields with infinitely many degrees of freedom 
describing massless modes, namely the electromagnetic (and the gravitational) field. For such 
systems one is able to solve the \textbf{``measurement problem''} of QM.  
(The states of individual isolated systems with finitely many degrees of freedom \textbf{not} coupled to 
any massless modes would actually evolve according to a linear, deterministic Schr\"odinger - von Neumann 
equation. For such (unphysical (!)) systems, actual events, or facts, would never emerge, and the 
``measurement problem'' cannot be solved.)}
\item{With regard to physical theory, a satisfactory completion of QM that describes the emergence 
of actual events, or facts, and solves the ``measurement problem'' appears to necessarily take 
the form of a \textbf{local} quantum theory in an even-dimensional space-time describing infinitely 
many degrees of freedom including \textbf{massless modes}, in particular photons. 

That the electromagnetic field apparently plays a fundamental role in a completion of QM 
may remind one of the fundamental role it has played in the genesis of the Special Theory 
of Relativity. This role of the electromagnetic field and of photons in completing QM has 
been missed in most discussions of the puzzles and problems of text-book QM we are
aware of.}
\item{We don't really know how Nature works deep down (the ``Ding an sich'' is not directly accessible 
to our senses and to our mind); but we should not stop striving for a unified description 
of natural processes manifested in experiments that is logically coherent and accounts for the 
observed facts in as precise a way as feasible. This is the underlying motto of our efforts.
}
\end{enumerate}

There have been previous attempts to extend conventional QM so as to obtain a satisfactory theory. 
Perhaps, the best known ones are Bohmian mechanics \cite{Durr} and spontaneous collapse theories 
\cite{GRW}; but there are other proposals, which we will mention. Our completion of QM shares some 
features with the ``Many Worlds Interpretation'' of QM, but amends this rather vague framework by a
precise one; and it describes only \textbf{one} world -- hopefully ours. All previous attempts to extend 
or complete text-book QM appear to suffer from some shortcomings. This is what has motivated us 
to look for yet another completion of QM; see \cite{FS}, \cite{BFS}, \cite{FP} and \cite{Detlef} 
for the original work. In this paper we present a sketch of the ideas and results described in these papers. 
Our results are based on novel, unfamiliar ideas and involve somewhat advanced mathematical concepts. 
This may explain why, for the time being, they have not caught much attention in the 
``quantum foundations community.'' With the exception of Section 6, the review offered in 
this paper only involves elementary mathematics, and we thus hope it will be widely understood 
and trigger discussion.

\subsection{Summary of contents}
The material contained in Section 1 is introductory; we argue rather carefully that there is a need to 
complete text-book QM in such a way that it can be used to describe the stochastic evolution of 
\textbf{individual} (isolated) physical systems, without invoking the action of ``observers'' who carry 
out ``measurements.'' In Section 2, we sketch an analogy between QM, on one hand, and the theory 
of diffusion processes and Brownian motion, on the other hand. Our presentation in Sections 1 and 2 
closely follows the one in \cite{Detlef}.

In Sections 3 and 4, we review the main ideas and results underlying the so-called 
\textbf{``\textit{ETH} - Approach to QM''}, where ``$E$'' stands for ``Events,'' ``$T$'' stands for ``Trees'' 
and ``$H$'' for ``Histories.'' The results described in these sections have appeared in several papers published 
in the course of the past ten years. 

Various applications of the $ETH$ - Approach to concrete phenomena, such as quantum-mechani- cal 
descriptions of measurements carried out on individual systems, the radioactive decay of nuclei, 
or the fluroescence of atoms coupled to the quantized electromagnetic field, etc. have been or will 
be presented elsewhere; but a short sketch of our recent results on fluorescence is included in 
Section 5. Section 6 contains a sketch of our approach to relativistic quantum theory. And 
Section 7 offers some conclusions.

\textbf{Remark:} Since this paper is a review of previously published results, it necessarily has 
much overlap with publications by some of the same authors, in particular with \cite{FS}, \cite{BFS}, \cite{FP} 
and \cite{Detlef}.

\subsection{Recap of text-book quantum mechanics}\label{text-book QM}

It may be appropriate to begin this paper with a description of some of the shortcomings of text-book 
QM and of its ``Copenhagen interpretation,'' with the intention to make it clear that there is an
obvious need for a \textbf{completion} of the theory. 

To bring this point home, once again, conventional (text-book) QM is a theory that only describes the 
behavior of \textbf{ensemble averages} taken over many identical physical systems and of the 
time evolution of ensemble-averaged states, henceforth called ``ensemble states.'' 
Ensemble states tend to be mixed states, their time evolution is \textbf{linear} and \textbf{deterministic}; 
it is goverened by the Schr\"odinger - von Neumann equation, or -- if the systems are in contact
with some \mbox{environment --} by a Lindblad equation. These equations do \textbf{not} describe 
the evolution of \textbf{individual} systems, which, for all we know, is \textbf{stochastic}. To see this 
one may consider the example of spontaneous emission of photons from excited atoms: 
the time and direction of emission of photons by an atom prepared in an excited state 
appear to be random variables. One would expect that a satisfactory completion of QM furnish 
the \textbf{laws} of such random variables and provide some fundamental stochastic 
equations describing the evolution of individual systems. 

Traditionally, the stochastic nature of the quantum-mechanical evolution of an individual system 
has been thought to be caused by ``measurements'' of physical quantities characteristic of the 
system in question; see \cite{Barchielli}, \cite{Gisin}, \cite{Diosi}, \cite{Belavkin}, \cite{BG}. 
However, as ought to become evident when studying a variety of examples of 
quantum-mechanical processes, such as fluorescence, the probabilistic/stochastic nature of the 
quantum-mechanical evolution of an individual physical system has, in general, little to do with
``measurements'' being carried out and even less with the intervention of ``observers'' and their 
consciousness, but, most often, with \textbf{``events''} released by the system that involve the emission 
of photons and possibly of other massless modes.  The formalism reviewed in Sections 3 and 4 takes 
these remarks into account and clarifies what ``events'' are in QM and in which way they are random.

There is a rather enlightening analogy between QM and the need to distinguish between 
individual systems and ensembles of identical systems, on one hand, and the theory of diffusion processes
and Brownian motion, on the other hand, which we will describe in some detail in Section 2: the evolution 
of ensembles of identical particles suspended in some liquid is described by a linear, deterministic equation, 
the diffusion equation, while an individual particle undergoes Brownian motion, which is genuinely stochastic 
and is described by the Wiener process. What we are looking for is a quantum-mechanical description 
of the stochastic evolution of the states of individual systems, i.e., for a quantum-mechanical analogue 
of the Wiener process; see Section 3.

We next recall the usual mathematical formalism used to describe text-book QM of ensembles 
of physical systems. Our presentation in the remainder of this section follows the one in \cite{Detlef}. 
This formalism is based on the following two pillars:
\begin{enumerate}
\item[(i)]{A \textbf{physical system}, $S$, is characterized by a list
\begin{equation}\label{O_S}
\mathcal{O}_{S}= \big\{\widehat{X}_{\iota}= \widehat{X}^{*}_{\iota} \big| \iota \in \mathfrak{I}_{S}\big\}
\end{equation}
of abstract bounded self-adjoint operators, where $\mathfrak{I}_{S}$ is a (continuous) set of indices. Every
operator $\widehat{X}\in \mathcal{O}_{S}$ represents a (bounded function of a) \textbf{physical quantity} 
characteristic of $S$, such as the electromagnetic field in a bounded region of space-time, or the 
total momentum, energy or spin of all particles (e.g., atoms) in $S$ localized in some bounded 
domain of physical space and interacting with the electromagnetic field. Of course, in QM, different 
operators in $\mathcal{O}_S$ do in general \textbf{not} commute with one another. One assumes that if 
$\widehat{X}\in \mathcal{O}_{S}$ and $F$ is a real-valued, bounded continuous function on $\mathbb{R}$ then 
$F(\widehat{X})\in \mathcal{O}_{S}$, too. In general $\mathcal{O}_{S}$ does not have 
any additional structure; (it is usually not a real linear space, let alone an algebra).

At every time $t$, there is a representation of $\mathcal{O}_{S}$ by bounded 
self-adjoint operators acting on a separable Hilbert space, $\mathcal{H}$,
\begin{equation}\label{1}
\mathcal{O}_{S} \ni \widehat{X} \mapsto X(t)=X(t)^{*} \in B(\mathcal{H})\,,
\end{equation}
where $B(\mathcal{H})$ is the algebra of all bounded operators on $\mathcal{H}$.

\textbf{Heisenberg-picture time evolution}: If $S$ is an \textbf{isolated} system, 
i.e., one whose interactions with the rest of the Universe are negligibly weak, then the operators 
$X(t)$ and $X(t')$ representing a physical quantity $\widehat{X}\in \mathcal{O}_{S}$ 
at two different times, $t$ and $t'$, are unitarily conjugated to one another. 
In an \textbf{autonomous} system,
\begin{equation}\label{2}
X(t')= e^{i(t'-t)H_S/\hbar} \,X(t)\,e^{-i(t'-t)H_S/\hbar}\,,
\end{equation}
where $H_S$ is the Hamiltonian of $S$. (For simplicity, we will henceforth usually
assume that $S$ is autonomous.) }
\item[(ii)]{``States,'' $\omega$, of $S$ are assumed to be given by density matrices, $\Omega$, i.e., 
by non-negative trace-class operators on $\mathcal{H}$ of trace unity. The expectation at time $t$ 
of an operator $\widehat{X}\in \mathcal{O}_{S}$ in a ``state'' $\omega$ of $S$ is given by
$$\omega\big(X(t)\big):=\text{Tr}\big(\Omega\,X(t)\big).$$
The state given by a density matrix $\Omega$ is said to be pure iff $\Omega$ is a rank-1 
orthogonal projection, $P=P^{*}=P^{2}$; otherwise it is called mixed.}
\end{enumerate}

\textbf{Remark}: The characterization of a physical system sketched above is \textbf{not} specific to QM. 
Classical Hamiltonian systems, too, are defined by specifying a complete list, $\mathcal{O}_S$, 
of physical quantities characteristic of $S$. In contrast to QM, if $S$ is a classical system the elements 
of $\mathcal{O}_S$ all commute with one another. This allows one to pass to the commutative algebra, 
$\mathcal{A}_S$, over the field of real numbers generated by $\mathcal{O}_S$, taken to be closed 
in the sup-norm ($L_{\infty}$-norm). The spectrum, $\Gamma_S$, of $\mathcal{A}_S$ is the 
\textbf{phase space} of the system, and $\mathcal{A}_S$ is identical to the algebra of real-valued
continuous functions, $C(\Gamma_S)$, over the phase space $\Gamma_S$.

The classical analogue of the Hamiltonian $H_S$ is a Hamiltonian vector field on $\Gamma_S$; and
states are given by probability measures on $\Gamma_S$. (It would be tempting to describe 
the analogy between QM and classical mechanics in more detail; but let's not.)

In text-book QM, it is usually assumed, following \textit{Schr\"odinger,} that, in the Heisenberg picture, 
``states'' of an isolated physical system $S$ are \textbf{independent} of time $t$, and, hence, that the 
Heisenberg picture is equivalent to the \textbf{Schr\"odinger picture}; namely
$$\omega(X(t))= \text{Tr}\big(\Omega\, X(t)\big)= \text{Tr}\big(\Omega(t)\, X\big), \quad X:=X(t_0),\, \Omega: =\Omega(t_0),$$
where $t_0$ is an (arbitrarily chosen) initial time. In the Schr\"odinger picture, 
the Schr\"odinger - von Neumann equation
\begin{equation}\label{S-L}
\dot{\Omega}(t) = -\frac{i}{\hbar}\big[H_S, \Omega(t)\big]\,, \quad t\in \mathbb{R}.
\end{equation}
describes the time evolution of states of $S$, while physical quantities of $S$ 
are represented by \textbf{time-independent} bounded self-adjoint operators, $X$, on $\mathcal{H}$. 

More generally, in the Schr\"odinger picture, the time-dependence of states of a system $S$ 
interacting with some environment is described by \textbf{linear, deterministic, trace-preserving, 
completely-positive maps}, $\big\{\Gamma(t, t')\big| t\geq t' \big\}$, 
\begin{equation}\label{Kraus}
\Omega(t)= \Gamma(t,t') \big[\Omega(t')\big]\,, \qquad \forall t\geq t'\,,
\end{equation}
where the operators $\Gamma(t,t')$ are defined on the linear space of trace-class operators on $\mathcal{H}$,
and $\Gamma(t,t')= \Gamma(t,t'')\cdot\Gamma(t'',t'), t\geq t'' \geq t',$ with $\Gamma(t,t)= \mathbf{1}$; see
\cite{Gorini, Lindblad, Kraus}.

While the formalism described up to here is adequate for a quantum-mechanical description of 
\textbf{large ensembles of identical systems} and, in particular, of the evolution of ensemble states,
it falls short of describing the behavior of \textbf{individual} systems and of the stochastic 
evolution of their states, as we will argue next.

\subsection{The inadequacy of text-book quantum mechanics}\label{inadequacy}
As already remarked several times, in text-book QM, the time evolution of ensemble states in 
the Schr\"odinger picture (see Eqs.~\eqref{S-L}, \eqref{Kraus}) is linear and deterministic. But, 
clearly, this cannot be the full story. As recognized by \textit{Einstein} in 1916 in his paper \cite{Einstein} 
on spontaneous and induced emission and absorbtion of light by atoms, which he described in probabilistic terms (introducing his $A$- and $B$-coefficients), QM is a \textbf{fundamentally probabilistic theory.} 
Text-book QM does not do justice to this insight and falls short of specifying a general probabilistic 
law enabling one to describe the stochastic evolution of individual physical systems. This paper 
reviews our attempts to fill this gap.

According to the Copenhagen interpretation of $QM$, the evolution of the state of an individual isolated
systems is described by the linear, deterministic Schr\"odinger - von Neumann equation -- 
\textbf{except} when an ``event'' happens, such as the emission or absorption of a photon by an atom, 
or the successful measurement of the value of a physical quantity $\widehat{X}\in \mathcal{O}_{S}$.
In this eventuality the Schr\"odinger - von Neumann evolution is ``interrupted.'' If the value of 
$\widehat{X}$ is measured at some time $t$, then the state of $S$ is claimed to make 
a \textbf{``quantum jump''} to a state in the range of the spectral projection of the operator $X(t)$ representing
$\widehat{X}$ corresponding to the value of $\widehat{X}$ measured at time $t$, i.e., 
corresponding to the eigenvalue of $X(t)$ associated with the measured value of $\widehat{X}$. 
It is claimed that QM predicts the probabilities or \textbf{frequencies} of ``quantum jumps'' to 
eigenstates corresponding to different possible values of $\widehat{X}$ when measurements 
of the value of $\widehat{X}$ are repeated many times for identical, identically prepared systems. 
These frequencies are supposed to be given by the \textbf{Born Rule} applied to the state of $S$ 
at the time when the measurement of $\widehat{X}$ begins. This summarizes (in rough terms) the contents 
of L\"uders' measurement postulate \cite{Luders}.

If the equipment used to measure the value of $\widehat{X}$ is included in what constitutes the total 
system $S$ -- now assumed to be isolated -- one might expect, erroneously, that the event corresponding 
to a measurement of the value of $\widehat{X} \in \mathcal{O}_S$ could be viewed as the result of the 
Schr\"odinger - von Neumann evolution of the state of the \textbf{total} system. This would appear to imply 
that QM is a deterministic theory -- \textbf{which it obviously isn't}, as remarked above; see \cite{Einstein}, 
and \cite{Wigner}, \cite{FFS} for more recent arguments. To repeat what we have already
said, the point is that the evolution of isolated physical systems capable of releasing events involving 
the emission of photons and, possibly, other massless modes turns out to be stochastic, 
and our task is to find the law describing such evolution.

It may be useful to take a brief look at an example, such as a Stern-Gerlach experiment, 
which is used to measure the vertical component of the spin ($s=\frac{1}{2}$) of a silver atom, 
exploiting what used to be called ``Richtungsquantelung.''
We imagine that the states of the spin of all silver atoms emitted by the atom gun indicated on the left of 
Figures 1 and 2 are prepared in the same way. For example, they may be given by the eigenstate of the 
$x$-component of the spin operator corresponding to the eigenvalue $+\frac{1}{2}$.
\begin{center}
\includegraphics[width=8cm]{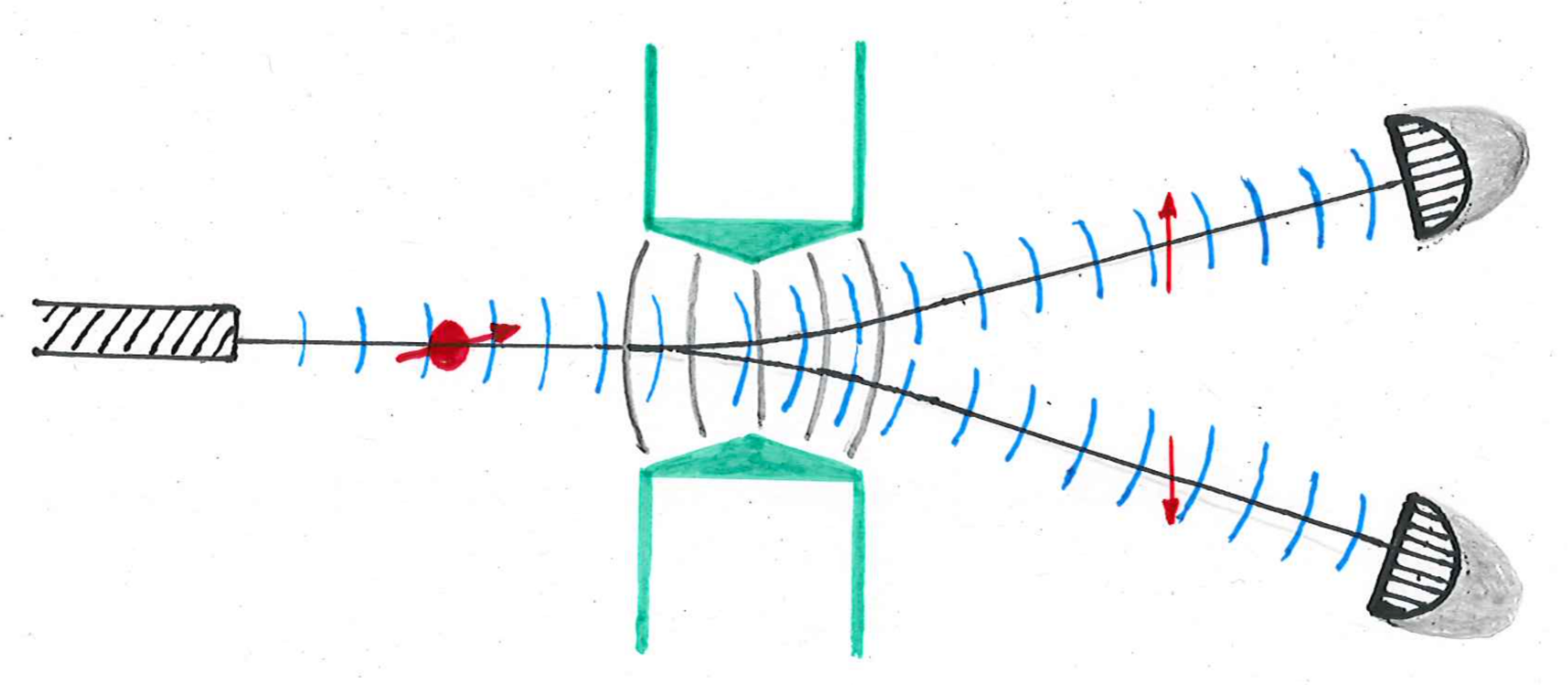}\\
{\small{Figure 1: \textit{Illustration of Schr\"odinger-von Neumann evolution}}}
\end{center}
Figure 1 is supposed to illustrate the \textbf{false} prediction derived from the Schr\"odinger - von Neumann 
equation for the ``state'' of the total system, including the detectors, that the orbital wave function of a 
silver atom evolves deterministically so as to always split into two pieces, one hitting the upper detector 
and the other one hitting the lower detector, which might then either both fire or remain mute.

Experiment shows, however, that every single silver atom only hits \textbf{one} detector, which then fires, 
(while the other one remains mute), as illustrated in \mbox{Figure 2.} The firing of the upper detector is 
interpreted as indicating that the state of the spin of the particular silver atom that has traveled through 
the device to hit the upper detector ends up being the eigenstate of the $z$-component of the spin 
operator corresponding to the eigenvalue $+\frac{1}{2}$, while the firing of the lower detector is 
taken to indicate that the state of the spin is in the eigenstate of the $z$-component of the 
spin operator corresponding to $-\frac{1}{2}$. In the process of the experiment, the states 
of the silver atom and of the detectors get entangled. Before the entanglement has taken 
place one cannot assign a specific trajectory to the silver atom.
\begin{center}
\includegraphics[width=8cm]{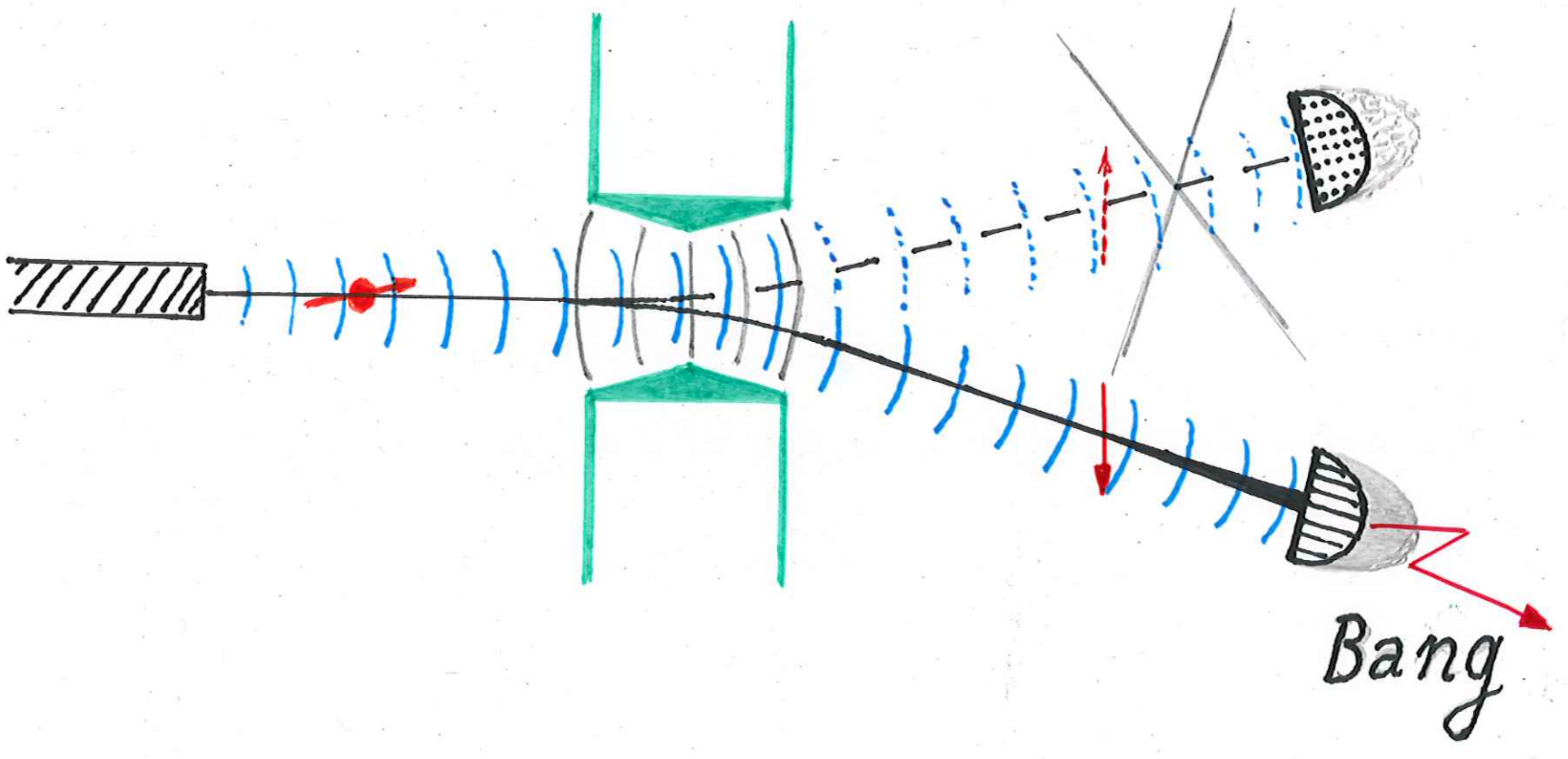}\\
{\small{Figure 2: \textit{Illustration of what people find in experiments}}}
\end{center}

The frequencies for the upper or the lower detector to fire when the same experiment
is repeated very many times are given by the usual Born Rule; (these frequencies would be given by 
$\frac{1}{2}$ if the initial state of the spin were an eigenstate of the $x$-component of the spin
operator). -- Apparently, experiment tells us that the evolution of the total system, which  includes the gun, 
an individual silver atom and the two detectors, is described by some \textbf{stochastic process}, 
\textbf{not} by a Schr\"odinger - von Neumann equation; the latter only describing the evolution of 
an \textbf{average} of states of very many identical, identically prepared systems, i.e., of an ensemble state.

A second example concerns the \textbf{fluorescence} of (i.e., spontaneous emission of photons 
from) a static, two-level atom coupled to the quantized electromagnetic field, which is put into a suitable 
laser beam that causes it to exhibit Rabi oscillations and, every once in a while, to spontaneously emit a photon. 
The two energy levels of the atom correspond to a ground-state and an excited state. It turns out 
that the spontaneous emission of photons is a stochastic process, which is accompanied by a 
``quantum jump'' of the atom to its ground-state, the times and directions of emission of the photons 
being random variables whose law one would like to determine. One would like to derive the stochastic 
process describing the fluorescence of such an atom from a proper completion of QM. A closely related 
example, which can be fully understood with the help of our completion of QM, will be considered in Section 5.\\

\textbf{Shortcomings of text-book QM}.
\begin{enumerate}
\item{\textbf{The ``measurement problem''}: The notion of ``events'' appears to be totally absent from 
the Copenhagen interpretation of QM and the notion of a ``measurement'' or ``observation'' remains
extremely vague. An example of a question apparently not answered by the Copenhagen 
interpretation is: What is the difference between a stage in the evolution of an isolated physical 
system without ``measurement,'' supposedly described by the Schr\"odinger equation, and a stage 
in its evolution when a ``measurement'' is carried out causing a ``quantum jump'' to occur, as suggested
in L\"uders' measurement postulate? To put it slightly differently, what is the 
precise criterion that enables one to decide whether, during a certain stage, the evolution 
of an isolated system is described by the Schr\"odinger equation or by some rule for wave function 
collapse?}
\item{\textbf{The problem of the role of ``observers''}: If one takes L\"uders' measurement postulate literally 
one is tempted to conclude that QM only makes useful predictions if it is known beforehand which 
measurements are planned by ``observers'' to be carried out, and at what times they will be carried out. 
One might then be misled to believe that something like the free will of ``observers'' plays a central role 
in working out testable predictions of QM about experiments and events.}
\item{\textbf{The problem of the infinitely short duration of measurements}: The hypotheses implicit 
in the Copenhagen interpretation that an ``observer'' can choose the time when a measurement 
begins, and that he or she can perform certain measurements that only take an arbitrarily small amount 
of time (which would actually imply that there are infinitely large energy fluctuations associated 
with such measurements) strike us as totally absurd.}
\item{\textbf{Randomness in the absence of ``observers'' and ``measurements''}: There are quantum phenomena, such as 
the radioactive decays of certain nuclei, in particular the precise decay times, or the fluorescence 
of atoms, that, for all we know, are intrinsically random, involving ``quantum jumps'' to states that,
fundamentally, cannot be predicted with certainty. There are no ``observers'' involved in triggering 
these jumps. So, where does the randomness of such phenomena originate from? Clearly they 
should be described by appropriate stochastic processes, which are not specified in text-book QM; 
and it is our task to identify and study these stochastic processes. }
\end{enumerate}

We do not expect that these shortcomings of text-book QM and its Copenhagen 
interpretation can be cured by some improved \textbf{``interpretation''} of what is an incomplete 
theory, among which one might mention ``Relational QM,'' ``QBism,'' ``Consistent Histories'' \cite{Griffiths}, 
\cite{G-M-H}, ``Many-Worlds Interpretation'' \cite{Everett}, ``Information ontologies,'' etc.; see \cite{Wiki} 
and references given there. We expect it to be equally unlikely that the shortcomings and paradoxes 
of text-book QM will be eliminated by supplementing this theory with some \textbf{ad-hoc mechanisms} 
that are supposed to explain what cannot be understood within the conventional theory, such as 
spontaneous wave-function collapse \cite{GRW}. These mechanisms remind us of electromagnetic 
or mechanical mechanisms that were advanced to explain Lorentz contraction before 
the advent of the Special Theory of Relativity. To us, attempts to reproduce the predictions of 
QM with the help of cellular automata \cite{Hooft}, to mention another example, look similarly unrealistic.

Bohmian mechanics is a logically coherent completion of (non-relativistic) quantum mechanics \cite{Durr}. 
But it reminds us of ``completing'' classical electrodynamics by introducing a mechanical medium, the ether, 
thought to be the carrier of electromagnetic waves. The Bohmian particles appear to be as unobservable 
as the ether. For, they are point-particles without any physical properties, such as electric charge or spin.
Bohmian mechanics does not appear to provide any laws that describe the precise effects of the Bohmian
particles when they hit detectors or screens, which then emit flashes of photons.
So, what is their role, why should one introduce them? (We know the standard objections against this
particular question, and we refrain from discussing them. We actually think the question is legitimate.) 
Furthermore, our (possibly wrong) impression is that a natural, ``canonical'' extension of Bohmian mechanics 
that includes relativistic quantum theory is not around the corner. It thus seems unlikely that Dirac would have 
regarded Bohmian mechanics or models of spontaneous wave-function collpase, etc. as acceptable 
completions of text-book QM -- we don't. (But see \cite{Bricmont} for some recent, vocal support of 
Bohmian mechanics.)

Before reviewing our own proposal, the $ETH$ - Approach, of how to complete QM in such a way 
that the theory makes precise statements about individual systems and their stochastic evolution, we 
sketch an analogy between quantum theory and the theory of diffusion processes and 
Brownian motion that is suggestive of what we are looking for.

\section{Analogy with the theory of diffusion processes}\label{BM}
To guide our thought process towards an adequate completion of QM with the help of an analogy, we briefly 
recall the theory of diffusion and Brownian motion. Consider a system consisting of a droplet, $\mathfrak{E}$,
of ink, e.g., eosin, in water (assumed to be in thermal equilibrium at some temperature $T>0$). 
The ``state'' of $\mathfrak{E}$ at time $t$ is given by its density, $\rho_t$, which is a non-negative 
function on physical space $\mathbb{E}^{3}$. We normalize it such that 
$\int_{\mathbb{E}^{3}} d^{3}x\, \rho_t(x) = 1$. The time dependence of $\rho_t$ is assumed 
to be governed by the diffusion equation, viz. by a \textbf{deterministic linear law} of evolution.
\begin{equation}\label{diffusion}
\dot{\rho}_{t}(x)= D\, (\Delta \rho_t)(x), \qquad D: \text{diffusion constant.}
\end{equation}
The well known solution of this equation is given by
\begin{equation*}
\rho_{t}(x)= \int_{\mathbb{E}^3} d^{3}x' \,\Gamma_{t-t'}(x-x') \rho_{t'}(x'), \quad \Gamma_{t}(x) := 
(2\pi Dt)^{-\frac{3}{2}} e^{-\frac{|x|^2}{2Dt}}\,,
\end{equation*}
and the ``heat kernels'' $\Gamma_t$ satisfy the \textit{Chapman-Kolmogorov} equation. 

According to the atomistic view of matter, $\mathfrak{E}$ really consists of very many eosin molecules, 
which, in an idealized description, can be viewed as tiny approximately spherical particles far separated 
from one another, so that interactions among these particles can be neglected, while they interact 
with the water molecules.
The state of an individual particle is its \textbf{position} in physical space. The ``state'' of $\mathfrak{E}$, 
given by its density $\rho$, should then be interpreted as an \textbf{ensemble average} over the states 
of the particles constituting the ensemble $\mathfrak{E}$ (i.e., as an ``ensemble state''). An individual 
system in this ensemble consists of a single particle. 
According to \textit{Einstein} and \textit{Smoluchowski} (1905), the particles in $\mathfrak{E}$ exhibit 
\textbf{Brownian motion} arising from random collisions with lumps of water molecules. From this 
assumption one can derive, for example, a formula for the diffusion constant (namely $D=\frac{k_B T}{6\pi \eta r}$).
We have learned from Einstein, Smoluchowski and \textit{Wiener} that Brownian motion ``unravels''
the diffusion equation, with the following \textbf{ontology}. 
\begin{enumerate}
\item[(i)]{ At every time $t$, the particle is located in some point $x_{\xi}(t)\in \mathbb{E}^{3}$ (its
initial state at some time $t_0$ being a point $x_0 \in \mathbb{E}^{3}$).}
\item[(ii)]{Its trajectory $\xi:=\big\{x_{\xi}(t)\big\}_{t\geq t_0}$ is a random continuous curve -- a Brownian path -- in 
physical space $\mathbb{E}^{3}$; but the velocity of the particle is ill-defined at all times.}
\item[(iii)]{Energy is conserved only for an ensemble of very many particles in thermal equilibrium with
the water, but \textbf{not} for an individual particle.}
\item[(iv)]{As shown by \textit{Wiener}, there exists a \textbf{probability measure}, $dW_{x_0}(\xi)$, 
on the space, $\Xi$, of particle trajectories, 
$\xi:= \big\{x_{\xi}(t)\in \mathbb{E}^{3}\,\big|\, t \geq t_0, \, x_{\xi}(t_0)=x_0 \big\}$, 
starting from $x_0$ at time $t_0$; this measure is supported on trajectories $\xi$ that are 
H\"older continuous of index $\frac{1}{2}$, etc.}
\item[(v)]{An ``event'' at time $t$ is the manifestation of the position, $x_{\xi}(t)$, of the particle. 
The trajectory $\xi$ can thus be viewed as a ``\textbf{history of events},'' a random object, 
and $\Xi$ is the ``space of histories.'' The manifestation of the position of a particle at some time
does not affect its future trajectory after that time. (In this respect, QM is of course different.)}
\end{enumerate}
Wiener measure $dW_{x_0}(\xi)$ allows us to predict probabilities of measurable sets of histories;
for example,
\begin{align}\label{LSW}
\begin{split}
\text{prob}\big\{\xi\in \Xi\,\big| x_{\xi}(t_i)\in &\mathcal{O}_i, \,i=1,2,\dots,n, \,t_0<t_1<\cdots<t_n\big\}\\
&=\int_{\Xi} dW_{x_0}(\xi) \prod_{i=1}^{n} \chi_{\{x_{\xi}(t_i)\in \mathcal{O}_i\}}\big(\xi\big)\,,
\end{split}
\end{align}
where $\mathcal{O}_1, \dots, \mathcal{O}_n$ are open sets in physical space, and $\chi_{\Delta}$ 
is the characteristic function of the set $\Delta \subset \Xi$.

The Chapman-Kolmogorov equation satisfied by the heat kernels implies that if regions 
$\mathcal{O}_{i}^{(\alpha)},$ $\alpha = 1,\dots, N$, for some $N$, are chosen such that
$\bigcup_{\alpha=1}^{N} \mathcal{O}_{i}^{(\alpha)}= \mathbb{E}^{3}$ then
\begin{align}\label{Markov}
\sum_{\alpha=1}^{N} \text{prob}\big\{\xi\,\big|& x_{\xi}(t_1)\in \mathcal{O}_1, \dots, x_{\xi}(t_i)\in \mathcal{O}_{i}^{(\alpha)},
\dots x_{\xi}(t_n)\in \mathcal{O}_n\big\}\nonumber\\
=&\text{prob}\big\{\xi\,\big|\, x_{\xi}(t_1)\in \mathcal{O}_1, \dots, x_{\xi}(t_{i-1})\in \mathcal{O}_{i-1}, x_{\xi}(t_{i+1})\in \mathcal{O}_{i+1}, \dots x_{\xi}(t_n)\in \mathcal{O}_n\big\}\,.
\end{align}
This property implies that if the position of a particle were measured at some intermediate time $t_i$ and 
then a sum over all possible outcomes of this measurement were taken one would obtain the \textbf{same} 
predictions for the outcomes of measurements of the particle positions at times earlier than $t_i$ and at 
times later than $t_i$ as if no measurement had been made at time $t_i$. This means that the retrieval of 
information about the state (position) of a particle does not affect its evolution. QM yields a different 
picture of reality: A measurement \textbf{always} affects the evolution of a \mbox{system --} even if the result 
of the measurement is \textbf{not} recorded. For this reason, an analogue of Equation \eqref{Markov} 
cannot generally hold in QM.

Using Wiener measure to take an average over the ensemble $\mathfrak{E}$ of very many identical 
particles, one recovers the \textbf{deterministic law} in Eq.~\eqref{diffusion} for the evolution 
of the ``state'' $\rho_t$,
\begin{align}\label{average}
\begin{split}
\int_{\mathcal{O}}d^{3}x\,\rho_{t}(x)= &\int_{\mathcal{O}}d^{3}x\,\int d^{3}x_0 \,\Gamma_{t-t_0}(x-x_0) \rho_{t_0}(x_0)\\
=& \int d^{3}x_0\, \rho_{t_0}(x_0) \int_{\Xi} dW_{x_0}(\xi)\, \chi_{\{x_{\xi}(t)\in \mathcal{O}\}}\big(\xi\big),
\end{split}
\end{align}
for an arbitrary open subset $\mathcal{O}\subset \mathbb{E}^{3}$.

We note that the Chapman-Kolmogorov equation for the heat kernels implies the 
Markov property for the Wiener measure $dW_{x_0}$, i.e., that a measurement 
of the particle position at some time $t$ wipes out all memory of its trajectory at times earlier
than $t$. In contrast, in quantum mechanics there usually are memory effects.
A quantum-mechanical analogue of the magic formula \eqref{LSW} for Brownian motion
has been proposed by \textit{L\"uders, Schwinger} and \textit{Wigner} (see \cite{Luders, S, W}). 
However, when applied to time-ordered series of measurements, 
their formula fails to satisfy an analogue of Eq.~\eqref{Markov}, because, in QM, the non-commutativity 
of operators representing different potential events that actualize at different times leads to \textbf{interference effects}. 
Not surprisingly, this has been noticed by many people, who thought of various ways to rescue 
a quantum-mechanical analogue of formula \eqref{LSW}. One formalism seemingly enabling 
one to rescue it, which has become quite popular, is known under the name of \textbf{``consistent histories''} 
\cite{Griffiths, G-M-H}. However, in our opinion, this formalism does \textbf{not} represent an 
acceptable completion of QM, because it talks about unpredictable and instantaneous 
interventions by ``observers'' who perform measurements, a feature that extinguishes much 
of the predictive power of QM. 

One might say that the Wiener measure \textbf{``unravels''} the diffusion equation \eqref{diffusion}.
In the next section, we describe an ``unraveling'' of the linear, deterministic Schr\"odinger - von Neumann 
evolution of ensemble states of identical systems by a non-linear, stochastic evolution of states of 
\textbf{individual systems} inspired by the observations concerning diffusion and Brownian motion 
just sketched. This will yield a satisfactory completion of QM and equip it with a plausible ``ontology.''

\section{The stochastic time evolution of individual systems}\label{ETH approach}
To set the stage for this section it may be appropriate to quote Werner \textit{Heisenberg},
who said: \textit{``Every experiment destroys some of the knowledge of the system which was obtained 
by previous experiments,''} which implies in particular that the initial state of a system does not 
precisely determine its future states. We will formulate a basic postulate, Axiom CP of Subsection 3.3, 
that gives Heisenberg's insight precise meaning.

The main purpose of the following considerations is to describe the \textbf{third pillar} to be added to the 
two conventional pillars of text-book quantum mechanics described in Sect.~1, in order to arrive at 
a complete theory. The \textbf{ontology} of our completion of QM resides in \textbf{``random 
histories of events''} (defined appropriately), in analogy to histories of positions (Brownian paths) 
occupied by a point-like particle exhibiting Brownian motion. In QM, we will attempt to equip the 
(non-commutative) space of histories of events with a \textbf{``(quantum) history measure''}; in 
analogy with the Wiener measure of Brownian motion. 
Our task is to find an appropriate notion of states of physical systems, to describe their non-linear 
stochastic time evolution and to introduce a quantum-mechanical analogue of the Wiener measure. 
The ETH - Approach to QM, developed during the past decade (see \cite{FS, BFS, Froh, FP}), 
accomplishes this task. Since, apparently, our proposal for how to complete QM is not widely 
known and appreciated, yet, we have to, once again, sketch it (what else can we do?). We follow 
the presentation in \cite{FZ}. In this section, we consider \textbf{non-relativistic QM}; a sketch 
of how to extend the approach summarized below to relativistic quantum theory is presented 
in Section 6.

\subsection{The \textit{ETH} - Approach to quantum mechanics}
Henceforth our discussion will employ the Heisenberg picture; and we consider \textbf{isolated} 
systems, i.e., systems, $S$, that have negligibly weak interactions with the rest of the Universe. 
For, only for isolated systems, the time-evolution of operators representing physical quantities 
of $S$ has a conceptually clear description in the form of the \textbf{Heisenberg equations of motion}. 
The main ingredients of the $ETH$ - Approach to the non-relativistic quantum theory of isolated systems
are the following ones.
\begin{enumerate}
\item[(I)]{As indicated in Section 1 (see Equation \eqref{O_S} of Subsection 1.1), we define a 
physical system $S$ by specifiyng a complete list of physical quantities
\begin{equation}
\mathcal{O}_{S}= \big\{\widehat{X}_{\iota}= \widehat{X}^{*}_{\iota} \big| \iota \in \mathfrak{I}_{S}\big\}\nonumber
\end{equation}
characterizing $S$, with properties \eqref{1}, \eqref{2} formulated in Subsection 1.1. 
Given $\mathcal{O}_S$, we may then introduce algebras \mbox{$\mathcal{E}_{\geq t}, t\in \mathbb{R},$} 
to be the weakly closed algebras (i.e., so-called von Neumann algebras) generated by all the operators
\begin{equation}\label{algebra}
\big\{\,X(t')\,\big|\, t'\geq t,\,\, \widehat{X} \in \mathcal{O}_S\,\big\}\,,
\end{equation}
where $t\in \mathbb{R}$ is time. 
Evidently, 
$$\mathcal{E}_{\geq t'}\, \subseteqq\,\, \mathcal{E}_{\geq t} \subseteqq B(\mathcal{H}), \,\text{ for }\,\, t'>t\,.$$

It is assumed that the two algebras $\mathcal{E}_{\geq t}$ and $\mathcal{E}_{\geq t'}$ are isomorphic 
to one another, for an arbitrary pair $(t,t')$ of times. For an \textbf{autonomous system} this follows from
the fact that these algebras are unitarily conjugated to one another:
\begin{align}\label{inclusion}
\mathcal{E}_{\geq t'}\, =\,\, e^{i(t'-t)H_S/\hbar}\, \mathcal{E}_{\geq t}\, e^{-i(t'-t)H_S/\hbar},\,\, \text{ for }\,\, t, t' \,\text{ in } \, \mathbb{R}\,,
\end{align}
where $H_S$ is the Hamiltonian of the system $S$. 
}
\item[(II)]{The notion of \textbf{``events''} -- in approximately the sense the late Rudolf \textit{Haag} must 
have used this terminology (see \cite{Haag}) -- plays a central role in the $ETH$-Approach: 
A \textbf{potential event} (or future potentiality) of $S$ setting in at a future time $\geq t$ is 
described by a partition of unity,
\begin{equation}\label{p-o-u}
\mathfrak{P}:=\big\{\pi_{\xi}\,\big|\, \xi \in \mathfrak{X}\big\} \subset \mathcal{E}_{\geq t}, 
\end{equation}
by orthogonal, mutually disjoint projections, $\pi_{\xi}$, with the properties that
\begin{equation}\label{projections}
\pi_{\xi}= \pi_{\xi}^{*}, \quad \pi_{\xi}\cdot \pi_{\eta}= \delta_{\xi \eta} \pi_{\xi},\,\,\, \forall\, \xi, \eta \in \mathfrak{X}, 
\quad \sum_{\xi \in \mathfrak{X}} \pi_{\xi}= \mathbf{1}\,,
\end{equation}
where $\mathfrak{X}$ is a finite or countably infinite set of labels called the \textbf{spectrum} of the potential
event $\mathfrak{P}$ and denoted by $\frak{X}= \text{spec}(\mathfrak{P}$). (In older, but misleading jargon, 
one may say that a potential event $\mathfrak{P}$ specifies a ``measurement basis.'') 
Given this notion of potential events, one may define the algebra $\mathcal{E}_{\geq t}$ as the 
(weak closure) of the algebra generated by all potential events in $S$ setting in at a future time $\geq t$. 
If $\mathfrak{P}$ is the potential event given by
$\mathfrak{P}=\big\{\pi_{\xi}\,\big|\, \xi \in \mathfrak{X}\big\} \subset \mathcal{E}_{\geq t},$ and defining
$$\pi_{\xi}(s):= e^{isH_S}\pi_{\xi}e^{-isH_S}, \quad \forall\,\,\xi \in \mathfrak{X}\equiv \text{spec}(\mathfrak{P}),$$
then 
$$\mathfrak{P}_s := \big\{\pi_{\xi}(s)\,\big|\, \xi\in \text{spec}(\mathfrak{P})\big\} \subset \mathcal{E}_{\geq t+s}$$
is a potential event setting in at a time $\geq t+s$.

The assumptions specified in ingredient (I), and in particular Eq.~\eqref{inclusion}, imply that all the algebras 
$\mathcal{E}_{\geq t}, t\in \mathbb{R},$ are isomorphic to \textbf{one} abstract weakly closed $^{*}$-algebra, 
denoted by $\mathcal{E}_{+}$. 

\textbf{Remark}: These observations enable one to define an isolated physical system by specifying an 
abstract (so-called $W^{*}$-) algebra, $\mathcal{E}_{+}$, generated by potential events, which we call 
\textbf{``algebra of future potentialities''}, and a family of isomorphisms, $\big\{\tau_{t}\,\big|\,t \in \mathbb{R}\big\}$,
\begin{align}\label{iso}
\begin{split}
&\tau_{t}: \mathcal{E}_{+} \rightarrow \mathcal{E}_{\geq t} \subseteq B(\mathcal{H}), \quad \text{with} \\
&\mathcal{E}_{\geq t'}=\tau_{t'}\circ \tau_{t}^{-1}\big( \mathcal{E}_{\geq t}\big) \subseteq \mathcal{E}_{\geq t}, \,
\text{ for }\, t'>t,
\end{split}
\end{align}
and, for isolated autonomous systems, $\mathcal{E}_{\geq t'}$ and $\mathcal{E}_{\geq t}$ are 
unitarily conjugated to one another, i.e., $\tau_{t'} \circ \tau_t$ in \eqref{iso} is given by unitary conjugation,
as specified in Equation \eqref{inclusion}. 

A generalized form of this remark is of importance in our approach to relativistic quantum theory (see
\cite{Fr}).
}
\item[(III)]{A \textbf{state} of an isolated system $S$ \textbf{at time} $t$ is given by a \textbf{quantum probability measure} 
on the lattice of orthogonal projections in $\mathcal{E}_{\geq t}$, i.e., by a functional, $\omega_t$, 
with the properties that
\begin{enumerate}
\item[(i)]{$\omega_t$ assigns to every orthogonal projection $\pi \in \mathcal{E}_{\geq t}$ a non-negative number 
$\omega_{t}(\pi) \in [0,1]$, \hspace{0.4cm} with $\omega_t(0)=0,$ and $\omega_{t}(\mathbf{1}) =1$; and}
\item[(ii)]{$\omega_{t}$ is additive, i.e.,
\begin{equation} \label{additivity}
\sum_{\pi \in \mathfrak{P}} \omega_{t}(\pi) = 1, \quad \forall\, \text{ potential events }\, 
\mathfrak{P} \subset \mathcal{E}_{\geq t}\,.
\end{equation}
A generalization of \textit{Gleason's} theorem \cite{GM} due to \textit{Maeda} \cite{Maeda} implies 
that states, $\omega_t$, of $S$ at time $t$, as defined above, are positive, normal, normalized linear 
functionals on $\mathcal{E}_{\geq t}$, i.e., states on $\mathcal{E}_{\geq t}$ in the usual sense of this notion,  
as employed in the mathematical literature. (Ignoring some mathematical subtleties) we henceforth identify 
$\omega_t$ with a density matrix on $\mathcal{H}$ denoted by $\Omega_t$.
}
\end{enumerate}
}
\item[(IV)]{An isolated \textbf{open} physical system, $S$, i.e., an isolated system capable of releasing ``events,''
is described by a ``co-filtration,'' $\big\{\mathcal{E}_{\geq t}\,\vert\, t\in \mathbb{R}\big\}$, of weakly closed
algebras (contained in the algebra, $B(\mathcal{H})$, of all bounded operators on $\mathcal{H}$) 
that satisfy the following

\vspace{0.1cm}\noindent \textbf{Principle of Diminishing Potentialities (PDP):} In an isolated \textbf{open}
system $S$ featuring events the following strict inclusions hold
\begin{equation}\label{PDP}
\hspace{1cm}\mathcal{E}_{\geq t}\,\, \supsetneqq\,\, \mathcal{E}_{\geq t'}\,, \,\text{ for arbitrary }\,\, t'>t\,. \hspace{1.5cm}\square
\end{equation}

People tend to be perplexed when confronted with PDP, because they find it hard to believe that
PDP is compatible with the unitary Heisenberg dynamics of operators described in Eqs.~\eqref{2} and \eqref{inclusion}.
However, in a relativistic local quantum (field) theory over an even-dimensional, flat space-time 
involving massless modes (photons, gravitons), such as quantum electrodynamics, and for the
correct choice of the algebras $\mathcal{E}_{\geq t}, t\in \mathbb{R}$, 
PDP can be shown to be a consequence of \textbf{Huygens' Principle}, 
as formulated and proven in \cite{Buchholz} in the context of algebraic quantum field theory. 
In \cite{FP}, some concrete models, including models arising when the velocity of light tends to $\infty$, 
are shown to satisfy PDP.}
\end{enumerate}

\subsection{Consequences of the Principle of Diminishing Potentialities}
The Principle of Diminishing Potentialities (PDP), when combined with the phenomenon of 
\textbf{entanglement,} implies that even if the state $\omega_t$ of $S$ at time $t$ were a 
``pure'' state on the algebra $\mathcal{E}_{\geq t}$ its restriction to the algebra $\mathcal{E}_{\geq t'}$ 
must be expected to be ``mixed'' at a later time $t'>t$. This observation opens the possibility to introduce 
the notion of ``events \textbf{actualizing} at some given time'' (and, in the end, to solve the ``measurement
problem'').

In accordance with the ``Copenhagen interpretation'' of QM, one might expect that a 
potential event,
$\mathfrak{P}=\big\{ \pi_{\xi}\, \big|\, \xi \in \text{spec}(\mathfrak{P}) \big\} \subset \mathcal{E}_{\geq t}$, 
becomes \textbf{actual} (manifest) at some time $\geq t$ iff
\begin{equation}\label{incoherent sp}
\text{tr}(\Omega_{t}\,A) = \sum_{\xi \in \mathfrak{X}} \text{tr}(\pi_{\xi}\, \Omega_{t}\,\pi_{\xi} \, A),\quad 
\forall A \in \mathcal{E}_{\geq t}\,.
\end{equation}
where $\Omega_t$ is the density matrix representing the state, $\omega_t$, of $S$ at time $t$.
Notice that, off-diagonal elements do not appear on the right side of \eqref{incoherent sp}, 
which thus describes an \textbf{incoherent} superposition of states in the images of disjoint 
orthogonal projections, i.e., a ``mixture.''

This expectation is made precise as follows. Given a state $\omega_t$ on $\mathcal{E}_{\geq t}$, 
we define $\mathcal{C}(\omega_t)$ to be the subalgebra of $\mathcal{E}_{\geq t}$ 
generated by all projections belonging to all potential events 
$\mathfrak{P}\subset \mathcal{E}_{\geq t}$ for which Eq.~\eqref{incoherent sp} holds. 
Further, $\mathfrak{P}(\omega_t)$ is the \textbf{finest potential event} contained in $\mathcal{C}(\omega_t)$ 
with the property that all its elements commute with all operators in $\mathcal{C}(\omega_t)$. (In more 
technical jargon, $\mathfrak{P}(\omega_t)$ generates the \textbf{center}, $\mathcal{Z}(\omega_t)$, 
of the \textbf{centralizer} $\mathcal{C}(\omega_t)$ of the state $\omega_t$ on the algebra $\mathcal{E}_{\geq t}$.)

We say that the potential event $\mathfrak{P}(\omega_t)$ \textbf{actualizes} at some time 
$\geq t$ iff $\mathfrak{P}(\omega_t)$ contains at least two non-zero orthogonal projections, 
$\pi^{(1)}, \pi^{(2)}$, which are disjoint, i.e., $\pi^{(1)}\cdot \pi^{(2)} =0$, and have non-vanishing 
Born probabilities, i.e.,
$$0< \omega_{t}(\pi^{(i)}) = \text{tr}\big(\Omega_t\,\pi^{(i)}\big) < 1\,, \quad \text{ for  }\, i=1,2\,.$$
Equation \eqref{incoherent sp} then holds true for $\mathfrak{P}=\mathfrak{P}(\omega_t)$, and the sum on the
right side of \eqref{incoherent sp} contains at least two distinct non-vanishing terms. One then says that
$\mathfrak{P}(\omega_t)$ (or, equivalently, $\mathcal{Z}(\omega_t)$) is \textbf{non-trivial}.

\textbf{Remark}: One may expect that there are physical systems with states $\omega_t$ for which 
the center $\mathcal{Z}(\omega_t)$ of the centralizer $\mathcal{C}(\omega_t)$ has continuous spectrum. 
The analysis presented in this and the next subsection must then be modified. It actually can be modified to
apply to such systems. (However, the necessary changes introduce an element of indeterminacy 
in the statement of Axiom CP, as proposed in the next Subsection. In the interest of keeping 
this paper understandable and reasonably elementary we will not present this extension; 
it will appear in forthcoming work).

\subsection{The state-reduction postulate and the stochastic evolution of states}
The \textbf{law} describing the non-linear stochastic time evolution of states of an individual isolated 
open system $S$ unraveling the linear deterministic evolution of ensemble states given in Equations
\eqref{S-L} and \eqref{Kraus} is derived from a \textbf{state-reduction postulate} described next. 
(The postulate has a precise mathematical meaning as long as time is discrete; the passage to
continuous time still presents a challenge, except in various concrete examples.)

Let $\omega_t$ be the state of $S$ at time $t$. Let $dt$ denote a time step; ($dt$ is strictly positive 
if time is discrete; otherwise one would attempt to let $dt$ tend to 0 at the end of the following constructions). 
We define a state $\overline{\omega}_{t+dt}$ on the algebra 
$\mathcal{E}_{\geq t+ dt} \,\,(\subsetneqq \mathcal{E}_{\geq t})$ by restriction of $\omega_t$ to the algebra 
$\mathcal{E}_{\geq t+dt}$,
\[\overline{\omega}_{t+dt}:= \omega_{t}\big|_{\mathcal{E}_{\geq t+dt}}\,.\]
As a manifestation of PDP and \textbf{entanglement,} the algebra $\mathcal{C}(\overline{\omega}_{t+dt})$
can be expected to be non-trivial (i.e., $\not= \mathbb{C}\cdot \mathbf{1}$) in general. This does, of
course, \textbf{not} imply that the potential event $\mathfrak{P}(\overline{\omega}_{t+dt})$ 
actualizing at some time $\geq t+dt$ is non-trivial, too. But it is plausible 
that it will in general be non-trivial. (This is shown to be the case in a family of models studied in \cite{FP}.)

We are now prepared to formulate a basic postulate, called state-reduction-, or \textbf{collapse postulate} (CP), 
which, together with the Principle of Diminishing Potentialities, is a corner stone of our approach. \\

\noindent {\textbf{Axiom CP:}} \,\,Let
$$\mathfrak{P}(\overline{\omega}_{t+dt})=\big\{\pi_{\xi}\,|\,\xi\in 
\text{spec}\big(\mathfrak{P}(\overline{\omega}_{t+dt})\big)\big\}$$ 
be the potential event actualizing at a time $\geq t+dt$, given the state $\overline{\omega}_{t+dt}$ 
on $\mathcal{E}_{\geq t+dt}$. Then `Nature' replaces the state $\overline{\omega}_{t+dt}$ on 
$\mathcal{E}_{\geq t+dt}$ by a state $\omega_{t+dt}$ given by
\begin{equation}\label{state red}
\omega_{t+dt}(X) \equiv \omega_{t+dt, \pi}(X):= \overline{\omega}_{t+dt}(\pi)^{-1}\,\overline{\omega}_{t+dt}(\pi\,X\,\pi),\quad \forall\,\,X\in \mathcal{E}_{\geq t+dt},
\end{equation}
(represented by the density matrix $\Omega_{t+dt, \pi}:= \text{tr}(\overline{\Omega}_{t+dt}\,\pi)^{-1} \cdot \pi\,\,\overline{\Omega}_{t+dt}\,\pi$), for some\,$\pi \in \mathfrak{P}(\overline{\omega}_{t+dt})$,
with $\overline{\omega}_{t+dt}(\pi)\not= 0$. The probability, $\text{prob}_{t+dt}(\pi)$, 
for the state $\omega_{t+dt,\pi}, \pi \in \mathfrak{P}(\overline{\omega}_{t+dt}),$ 
to be selected by `Nature' as the state of $S$ at time $t+dt$ is given by \textbf{Born's Rule}
\begin{equation}\label{Born Rule}
\hspace{5cm}\text{prob}_{t+dt}(\pi)= \overline{\omega}_{t+dt}(\pi) = \text{tr}(\overline{\Omega}_{t+dt}\,\,\pi)\,. 
\hspace{3cm}\square
\end{equation}
Since the elements $\pi \in \mathfrak{P}(\overline{\omega}_{t+dt})$ do not necessarily commute with the
Hamiltonian, $H_S$, of the system, energy is generally \textbf{not} conserved in the passage from $\omega_t$
to $\omega_{t+dt}=\omega_{t+dt, \pi}$. (It is however conserved in the mean.)

The projection $\pi(t+dt):=\pi \in \mathfrak{P}(\overline{\omega}_{t+dt})$ appearing in \eqref{state red}
and \eqref{Born Rule} is called \textbf{actual event} or \textbf{actuality} (or fact) at time $t+ dt$.

The analogue of the initial position, $x_0$, of a Brownian path at time $t_0$ is the initial state 
$\omega_0$ on $\mathcal{E}_{\geq t_0}$; the analogue of the Brownian trajectory 
$\xi=\big\{x_{\xi}(t)\,\big|\,t\geq t_0\big\}$ is given by a \textbf{history},
$\mathfrak{h}:=\big\{\pi(t_0+dt), \pi(t_0 + 2dt), \dots, \pi(t)\big\}$, of \textbf{actual events} originating
from the initial state $\omega_0$ of $S$ at time $t_0$. With a history $\mathfrak{h}$ we 
associate a \textbf{``history operator''} defined by
$$H_{\mathfrak{h}}(t_0, t):= \prod_{t'\in \mathbb{Z}_{dt},\, t_0 < t' \leq t} \pi(t') \,.$$
In quantum mechanics, the role of the Wiener measure, $dW_{x_0}$, of Brownian motion is played
by the probabilities
\begin{align}\label{prob history}
\text{prob}_{\omega_{0}}\big[ \mathfrak{h}=&\big\{\pi(t_0+ dt), \pi(t_0 + 2dt),\dots, \pi(t)\big\}\big]:=\nonumber\\
&=\omega_{0}\big(H_{\mathfrak{h}}(t_0,t)\cdot H_{\mathfrak{h}}(t_0,t)^{*}\big)
=\text{tr}\big[H_{\mathfrak{h}}(t_0,t)^{*}\cdot \Omega_{0}\cdot H_{\mathfrak{h}}(t_0,t)\big]
\end{align}
of histories of events, where $\Omega_o$ is the density matrix representing the initial
state $\omega_0$ on the algebra $\mathcal{E}_{\geq t_0}$.

It follows from our discussion that the time-evolution of the state of an \textbf{individual} physical 
system $S$ is described by a \textbf{stochastic branching process}, called ``quantum Poisson process,''
whose ``state space'' is referred to as the \textbf{non-commutative spectrum}, $\mathfrak{Z}_{S}$, of $S$ 
and is defined as follows. By equation \eqref{inclusion}, all the algebras $\mathcal{E}_{\geq t}$ are isomorphic 
to one specific (universal) von Neumann algebra, which we have denote by $\mathcal{E}_{+}$. 
The non-commutative spectrum, $\mathfrak{Z}_{S}$, of $S$ is defined by
\begin{equation}\label{NCspect}
\mathfrak{Z}_{S}:= \bigcup_{\omega} \Big(\,\omega\,, \mathfrak{P}(\omega)\Big)\,, 
\end{equation}
where the union over $\omega$ is a disjoint union, and $\omega$ ranges over all states on 
$\mathcal{E}_{+}$ of physical interest. (``States of physical interest'' are normal states on 
$\mathcal{E}_{+}$ a system can actually be prepared in.) 
The branching rules of a quantum Poisson process on $\mathfrak{Z}_S$ are uniquely determined 
by \textbf{Axiom CP}.

\vspace{0.2cm}\textbf{Comments and Remarks}.
\begin{itemize}
\item{One may expect -- and this can be verified in concrete models (see \cite{FP} 
for further details) -- that, most of the time, the actual event, $\pi \in \mathfrak{P}(\overline{\omega}_{t+dt})$, 
which, according to the Born Rule, has the largest probability to happen, and hence is most likely 
to be chosen by `Nature' (see \eqref{state red}), has the property that
\begin{equation}\label{trivial evol}
\omega_{t+ dt}\equiv \omega_{t+dt, \pi} \approx \overline{\omega}_{t+ dt} = \omega_{t}\big|_{\mathcal{E}_{\geq t+dt}}\,.
\end{equation}
This would imply that, most of the time, the evolution of the state is close to being trivial (as assumed
in text-book QM in the absence of ``measurements''). But, every once in a while, the 
state of the system makes a \textbf{``quantum jump''} corresponding to an actual event $\pi$ in
\eqref{state red} that is very unlikely to materialize. Such ``quantum jumps'' happen for purely entropic
reasons at random times.}
\item{One may check that the non-linear stochastic evolution of states outlined above has the 
desirable feature that it reproduces the usual Schr\"odinger - von Neumann evolution when an ensemble-average 
over all possible histories of very many identical systems is taken.}
\item{Our construction of the non-linear stochastic time evolution of individual systems is meaningful, 
mathematically, as long as $dt >0$; but, for the time being, the limiting theory, as $dt \searrow 0$, 
is only understood precisely in examples (see Sect.~5).}
\item{A \textbf{passive state} is a state $\omega_t$ on the algebra $\mathcal{E}_{\geq t}$ of potentialities 
at times $\geq t$ with the property that $\overline{\omega}_{t'}:=\omega_t\big|_{\mathcal{E}_{\geq t'}}$ 
has a centralizer with a \textbf{trivial center} $\mathcal{Z}(\overline{\omega}_{t'})$ (as defined above), 
\textbf{for all} $t'>t$. It then follows that the states $\omega_{t'}$ occupied by the system at times 
$t'>t$ are given by the states $\omega_t\big|_{\mathcal{E}_{\geq t'}}$, for all $t'>t$.

States of isolated autonomous systems satisfying PDP may be expected to relax towards passive states,
as time tends to $\infty$. For example, the state of a system consisting of an isolated atom coupled to the
quantized electromagnetic field is expected to relax towards a state describing the atom in its ground state,
which looks like the vacuum state of the electromagnetic field far away from the atom, as time tends to $\infty$.

Ground states and infinite-volume limits of equilibrium states of systems with infinitely many degrees 
of freedom tend to be passive states.}
\item{The principle of Diminishing Potentialities (PDP), in combination with the state reduction postulate, i.e.,
Axiom CP of Subsection 3.1, is an expression of the fact that the completion of QM proposed here -- which 
describes events and solves the measurement problem -- features a fundamental \textbf{``Arrow of Time''}, 
as emphasized in \cite{Arrow}. Actually, the idea of an arrow of time in the evolution of Nature goes back 
to the ancient times of \textit{Aristotle}, who wrote: \textit{Indeed, it is evident that the mere passage 
of time itself is destructive rather than generative ..., because change is primarily a ‘passing away.’}}
\end{itemize}

In earlier work (see \cite{BFS}), PDP was called \textbf{``Principle of Loss of (Access to) Information''} 
(LAI). Indeed, PDP and Axiom CP imply that information about an individual physical system 
acquired in the past tends to get lost in the course of its stochastic evolution (as claimed by Heisenberg).
Moreover, autonomous systems satisfying PDP usually have the property that events become increasingly 
``rare,'' as time increases, because their states approach a passive state, as mentioned above. 
These features are concrete manifestations of the arrow of time built into the foundations of our 
completion of QM. They appear to enable one to solve (or rather dissolve) not only the measurement
problem but also the ``unitarity- or information paradox,'' to mention two examples of enigmatic
puzzles of text-book QM. While we have emphasized these and other successes of our 
completion of QM for the past decade, they appear to attract some attention only now (see, e.g., \cite{BF}).

\section{The Principle of Diminishing Potentialities as a consequence of Huygens' Principle}
Our aim, in this section, is to identify a concrete physical mechanism that implies the validity of the Principle 
of Diminishing Potentialities (PDP) in realistic models of isolated physical systems with matter degrees 
of freedom coupled to massless modes, photons and/or gravitons. Rather than reviewing the 
general theory, which is based on \cite{Buchholz}, we consider an example.\\

\noindent\textbf{Huygens' Principle in an idealized system:}
Let $S$ be an isolated system consisting of a static atom located near the origin, $\mathbf{x}=0$, in physical
space and coupled to the quantized electromagnetic field through an electric dipole moment. We assume that
\begin{itemize}
\item{the atom has $M$ energy levels; hence its Hilbert space of state vectors is given by
$\mathfrak{h}_A\simeq \mathbb{C}^{M}$, and, before it is coupled to the radiation field, its Hamiltonian 
is given by a symmetric $M\times M$ matrix,
$H_A$, acting on $\mathfrak{H}_A$;}
\item{the Hilbert space of the free electromagnetic  field is the usual Fock space, $\mathfrak{F}$, of photons.
The quantized electromagnetic field is described by its field tensor, $F_{\mu \nu}(t, \mathbf{x})$, with 
$x=(t, \mathbf{x})$ a point in Minkowski space-time $\mathbb{M}^4$, which is an operator-valued distribution 
with the property that, for real-valued test functions $\big\{h^{\mu \nu}\big\}$ on space-time, 
\[ F(h):= \int_{\mathbb{R}\times \mathbb{R}^{3}} dt\, d\mathbf{x} \, F_{\mu \nu}(t, \mathbf{x})\,h^{\mu\nu}(t, \mathbf{x})\]
is a self-adjoint operator on $\mathfrak{F}$ that satisfies \textbf{locality} in the form of ``Einstein causality.''

The energy-momentum operator, $(H_f, P_f)$, of the free electromagnetic field satisfies the usual 
relativistic \textbf{spectrum condition}, i.e., the spectrum of  $(H_f, P_f)$ is contained in the forward 
light cone; in particular the Hamiltonian $H_f$ is a non-negative operator on $\mathfrak{F}$. Fock space 
contains a vacuum vector, $\big| \emptyset\big>$, with the property that $H_f \big| \emptyset\big>=0, \,\,
P_f \big| \emptyset\big>=0$.}
\end{itemize}

Since the atom is located near $\mathbf{x}=0$ and is static, it is useful to introduce the space-time diamonds 
$$D_{[t,t']}:= V^{+}_{t} \cap V^{-}_{t'}, \,\,\, t'>t,$$ 
centered on the time axis ($\mathbf{x}=0$), with $V^{\pm}_{t}$ the forward or backward light cone, respectively, 
with apex in the point $(t, \mathbf{x}=0)$ of the time axis. We will also consider models arising when the speed
of light, $c$, tends to $\infty$. These models will serve to illustrate PDP in the context of non-relativistic 
QM; (see also \cite{FP}).

\subsection{A (not so) simple model}
The Hilbert space of the system $S$ is chosen to be
$$\mathcal{H}:=\mathfrak{h}_{A}\otimes \mathfrak{F}\,.$$
Bounded functions of the field operators $F(h)$, with $h^{\mu\nu}$ real-valued and supported in $D_{[t,t']}$, 
for all $\mu, \nu,$ generate a von Neumann algebra $\mathcal{A}_{I=[t,t']}$. We define the von Neumann 
algebras
\begin{align}\label{algebras}
\begin{split}
\mathcal{D}_{I}^{(0)} := \mathbf{1}\big|_{\mathfrak{h}_A}\otimes \mathcal{A}_I\,, & \qquad
\mathcal{E}_{I}^{(0)}:= B(\mathfrak{h}_{A})\otimes \mathcal{A}_{I}\,,\\
\mathcal{E}_{\geq t}^{(0)}:= &\overline{\bigvee_{I \subset [t,\infty)} \mathcal{E}_{I}^{(0)}}\,,
\end{split}
\end{align}
(where the closure is taken in the weak$^{*}$ topology). 

We first convince ourselves that PDP holds for this system \textbf{before} the atom is coupled to the 
electromagnetic field. The Hamiltonian of the system is then given by 
$H_0:= H_A \otimes \mathbf{1} + \mathbf{1} \otimes H_f$, and
$$\mathcal{E}^{(0)}_{\geq t'}= e^{i(t'-t)H_f} \mathcal{E}^{(0)}_{\geq t} e^{i(t-t')H_f}, \quad \forall\,\, (t',t).$$
If $\mathcal{M} \subseteq B(\mathcal{H})$ is an algebra of bounded operators acting on the Hilbert space 
$\mathcal{H}$ then $\mathcal{M}'$ denotes the algebra of all those bounded operators on $\mathcal{H}$ 
that commute with \textbf{all} operators in $\mathcal{M}$; the algebra $\mathcal{M}'$ is called the 
\textbf{commutant} of $\mathcal{M}$.\\
Before the atom is coupled to the electromagnetic field one has that
\begin{equation}\label{rel commutants}
\big[\mathcal{E}_{\geq t'}^{(0)}\big]' \cap \mathcal{E}_{\geq t}^{(0)} = \mathcal{D}_{[t,t']}^{(0)}\,,\quad \text{for arbitrary }\,\, t'>t.
\end{equation} 
Noticing that, for arbitrary $t'>t$,  $\mathcal{D}_{[t,t']}^{(0)}$ is an $\infty$-dimensional algebra, we find that 
\eqref{rel commutants} is a very strong form of PDP.

Property \eqref{rel commutants} follows from \textbf{``Huygens' Principle''} in a straightforward way, 
namely from
\begin{equation}\label{loc comm}
[F_{\mu\nu}(x), F_{\rho, \sigma}(y)]=0, \,\,\forall\, \mu, \nu, \rho, \sigma, \,\, \text{ unless }\,\,x-y \text{ is \textbf{light-like}},
\end{equation}
a fact that, for the free electromagnetic field, can be found in every decent book on quantum field theory.
That a property closely related to \eqref{rel commutants} also holds in relativistic quantum 
electrodynamics with non-trivial interactions is a deeper fact that has been established in \cite{Buchholz}. 

To come up with a model that has a precise mathematical meaning, we make use of an ultraviolet 
regularization of quantum electrodynamics arising from discretizing time, $t_{n} := n\,\tau,\,\, n \in \mathbb{Z},$ where\, 
$\tau>0$\, denotes the time step, \,(with $\tau \equiv dt$, in the notation of Sect.~3).
To describe interactions, we pick a unitary operator $U \in \mathcal{E}_{[0,\tau]}^{(0)}$ and define
$$U_{k}:= e^{i(k-1)\tau H_f} \,U\, e^{-i(k-1)\tau H_f}, \quad k=1,2,\dots, \quad \text{and }\,\,U(n):=\prod_{k=1}^{n} U_{k}\,.$$
\begin{equation}\label{propagator}
\Gamma:= e^{-i\tau H_f} U \,\, \Rightarrow\,\, \Gamma^{n}=e^{-in\tau H_f} U(n), \,\, (\Gamma^{n})^{*}=: \Gamma^{-n}, \,\, n=0,1,2,\dots,
\end{equation}
with $\Gamma^{0}=\mathbf{1}$. The operators $\big\{\Gamma^{n}\big\}_{n\in \mathbb{Z}}$ represent the 
propagator of an interacting system with discrete time.

To study the dynamics of this model it suffices to consider the time evolution for times $t \geq t_0:=0$. We define
\begin{equation}\label{interacting algebra}
\mathcal{E}:=\mathcal{E}_{\geq 0}^{(0)}, \quad \mathcal{E}_{\geq n}:= \big\{ \Gamma^{-n}\,X\,\Gamma^{n}\,\big|\,X\in 
\mathcal{E}\big\}\,.
\end{equation}

\textbf{PDP for the interacting model}: It is not difficult to show, using \eqref{propagator} and 
\eqref{interacting algebra}, that
\begin{equation}\label{PDP interacting}
\big[\mathcal{E}_{\geq n'}\big]' \cap \mathcal{E}_{\geq n} \simeq \mathcal{D}_{[n,n']}\,, \,\,\,\text{for }\,\,n'>n,
\end{equation}
where \,\,$\mathcal{D}_{[n,n']}:= \big\{U(n')^{*}\, X\, U(n')\,\big|\, X\in \mathcal{D}_{[n\tau,n'\tau]}^{(0)}\big\}$.\\

Preparing the interacting system $S$ in an initial state $\omega_0$ at time $n=0$, e.g., one where the 
electromagnetic field is in the vacuum state $\big|\emptyset \big>$ and the atom is in an excited state, 
one may attempt to describe the stochastic time evolution of the state of $S$, as determined by the 
$ETH$ - Approach, in particular by \textbf{Axiom CP} of Subsect.~3.3. This will yield a description of 
spontaneous emission of photons by the atom, that is fluorescence of atoms, as a ``quantum stochastic process.'' 
It is however rather difficult to come up with explicit results, because, as long as the velocity of light, $c$, is finite, 
the electromagnetic field gives rise to memory effects related to the fact that expectations of field operators localized 
in diamonds belonging to different time slices do not factorize in states of physical interest. (Put differently, memory 
effects are due to electromagnetic fields in different time slices not being statistically independent.) 
In order to avoid the difficulties caused by these memory effects, we adopt a non-relativistic description 
of the system $S$ that emerges in the limit where $c$ tends to $\infty$. 

\section{Fluorescence of two-level atoms coupled to the quantized radiation field}
In this section, we study \textbf{fluorescence} of very heavy two-level atoms coupled to the 
radiation field; (see, e.g., \cite{Pomeau} and refs.~given there). In order to be able to state explicit results, 
we study this phenomenon in the ``non-relativistic'' limit, $c\rightarrow \infty$, 
which drastically simplifies our analysis. In this limit, the light cones $V^{+}_t$ unfold to become planes
\mbox{$\big\{x= (t, \mathbf{x})\,\big| \, t \text{ fixed}, \,\mathbf{x}\in \mathbb{E}^{3}\big\}$,} and the 
space-time diamonds $D_{[t',t'']}$ introduced in the last section become time slices, 
$$D_{[t',t'']} \rightarrow \big\{(t, \mathbf{x})\,\big|\, t'\leq t < t'', \mathbf{x}\in \mathbb{E}^{3}\big\}.$$
The natural form of Huygens' Principle in the limit where $c\rightarrow \infty$ is to require that functionals 
of the ``radiation field'' localized in different time slices \textbf{commute}. The field Hamiltonian 
$H_f$ then becomes the generator, $P_0$, of translations in the direction of the time axis, and the 
algebras $\mathcal{D}_{[t,t+\tau]}^{(0)}$ are ``full matrix algebras,'' $\simeq B(\mathcal{H}_{\tau}),$ \,
where\, $\mathcal{H}_{\tau}$ is a separable Hilbert space (described below), for arbitrary $\tau>0$. If the initial 
state of the radiation field is chosen to be a ``product state'' factorizing over different time slices, concretely 
the vacuum state $\big|\emptyset \big>$ introduced below, then the time evolution of ensemble-averaged 
states (``ensemble states'') of $S$ becomes \textbf{``Markovian.''} It is described by a Lindblad equation
\cite{Lindblad}. Very explicit results can then be obtained. 

It may help the reader to first consider models with discrete time, as in Sections 3 and 4 and in \cite{FP}.

\subsection{An explicit model of fluorescence}

We imagine that, every $T$ seconds, an atom source releases a two-level atom prepared in a superposition of 
a ground state, $|\downarrow\rangle$, and an excited state, $|\uparrow\rangle$. In somewhat less than 
$T$ seconds, such an atom propagates to an atom-detector, where an ``observable,'' such as
$X:=\begin{pmatrix} 1&0\\0&-1 \end{pmatrix}$, acting on the Hilbert space, $\mathfrak{h}_A=\mathbb{C}^{2}$, 
of internal states of the atom, is measured; (for descriptions of measurements within the $ETH$ - Approach
to QM see \cite{FP}). During its trip from the atom source to the detector, an atom may jump from 
$|\uparrow\rangle$ to $|\downarrow\rangle$ and emit a ``photon,'' $\gamma$, as first 
studied by \textit{Einstein} in 1916. (Since we consider the model in the limit $c\rightarrow \infty$,
we put the expression ``photon'' in quotation marks.) Different atoms are treated as statistically 
independent -- there are no correlations between the states they are prepared in.\vspace{0.2cm}\\
We consider two different experiments.
\begin{enumerate}
\item{A ``photon'' possibly emitted by an atom on its trip from source to detector escapes the experimental 
setup and is \textbf{not} detected before any entanglement between its state and the state
of the atom is wiped out.}
\item{A ``photon'' possibly emitted by such an atom ``immediately'' hits a photo-multiplier that fires when hit
by the ``photon'' before the atom ends its trip to its own detector, where the ``observable'' $X$ is measured; 
entanglement between the state of the photon and the state of the atom is preserved until the two measurements
occur.}
\end{enumerate}
The Hilbert space of the total system is given by 
$$\mathcal{H}:=\mathfrak{h}_{A}\otimes \mathfrak{F} \otimes \mathfrak{H}_{\gamma}\,,$$
where $\mathfrak{h}_A=\mathbb{C}^{2}$ is the Hilbert space of a two-level atom (whose orbital degrees
of freedom can be ignored in the following), and $\mathfrak{F}$ is the 
Fock space of the ``radiation field'' (in the limit where $c\rightarrow \infty$), which is defined below. 
Moreover, $\mathfrak{H}_{\gamma}$ is the Hilbert space of the photo-multiplier; it will not enter 
the following considerations explicitly. In experiment 1, the photo-multiplier is turned off and can be 
entirely neglected.

It is convenient to parametrize the states of an atom by density matrices on 
$\mathfrak{h}_A$, $\rho(\vec{n})$, given by
\begin{equation}\label{Bloch param}
\rho(\vec{n}):= \frac{1}{2}\big(\mathbf{1}_2 + \vec{n}\cdot \vec{\sigma}\big),\quad \vec{n}\in \mathbb{R}^{3}, 
\text{ with } |\vec{n}\, | \leq 1\,,
\end{equation}
where $\vec{\sigma}:= \big(\sigma_1, \sigma_2, \sigma_3\big)$ is the vector of Pauli matrices. We recall that
the state $\rho(\vec{n})$ is \textit{pure}, i.e., $\rho(\vec{n})$ is a rank-1 orthogonal projector, iff $|\vec{n}\,|=1$
(i.e., iff $\vec{n}$ belongs to the ``Bloch sphere'').
Moreover,
$\rho(\vec{n}) + \rho(-\vec{n})=\mathbf{1}$, and $\text{tr}\big( \vec{\sigma} \cdot \rho(\vec{n})\big) = \vec{n}.$
The matrix $\rho(\vec{n})$ has eigenvalues $\frac{1+|\vec{n}|}{2}$ and $\frac{1-|\vec{n}|}{2}$, 
with eigenspaces given by the ranges of the projections $\rho(\pm \frac{\vec{n}}{|\vec{n}|})$, respectively.

The Hamiltonian, $H_A$, of an atom decoupled from the radiation field is given by
\begin{align}\label{Hamiltonian}
H_A:= (1/2)\vec{\omega}\cdot \vec{\sigma},\,\, \text{ with }\,\vec{\omega}= (0,0, \Omega).
\end{align}
One then has that
\begin{align}
e^{itH_A}\rho(\vec{n}_0)e^{-itH_A}=& \rho\big(\vec{n}(t)\big), \,\, \text{ where }\,\,
\vec{n}(t)= (\text{sin}\theta_0 \, \text{cos}\varphi(t), \text{sin}\theta_0 \, \text{sin}\varphi(t), \text{cos}\theta_0),
\end{align}
with $\varphi(t)= \varphi_0 + \Omega\cdot t$ and   $\vec{n}_0 = \vec{n}(t=0)=
(\text{sin}\theta_0 \, \text{cos}\varphi_0, \text{sin}\theta_0 \, \text{sin}\varphi_0, \text{cos}\theta_0)$; ($(\theta, \varphi)$ 
are the usual polar angles).

The \textbf{Fock space,} $\mathfrak{F}$, of the radiation field is defined as follows. We introduce creation- 
and annihilation operators $a^{*}(t, X)$ and $a(t, X), \text{ with }\, t \in \mathbb{R},\, X \in \mathcal{X}$, 
where $\mathcal{X}$ represents ``physical space'' (which, for simplicity, we may suppose to be a finite set of points) and
the possible polarizations of a ``photon.'' The operators $a^{*}(t, X)$ and $a(t, X)$ satisfy the canonical commutation relations
\begin{align}\label{CCR}
[a(t, X), a^{*}(t', X')]= \delta(t-t')\cdot C_{X\,X'} \qquad 
[a^{\#}(t, X), a^{\#}(t', X')] = 0\,,
\end{align}
for all $t, t'$ in $\mathbb{R}$ and all $X, X'$ in $\mathcal{X}$, where $a^{\#} = a \text{ or } a^{*}$,
and $\big\{C_{X\,X'}\big| X, X' \text{ in } \mathcal{X}\big\}$ are the matrix elements of a bounded, 
positive-definite quadratic form $C$. We introduce a ``vacuum vector,'' $\big|\emptyset\big>,$ with the properties that 
\begin{equation}\label{vacuum}
a(t, X)\big|\emptyset \big> = 0, \quad \forall t\in \mathbb{R},\,\, \forall X \in \mathcal{X}\,, \quad \text{and }\,\,
\big< \emptyset \big| \emptyset\big> =1.
\end{equation}
Fock space $\mathfrak{F}$ is defined to be the completion in the norm induced by the scalar product 
\mbox{$\big<\cdot \big| \cdot\big>$} (uniquely determined by \eqref{CCR} and \eqref{vacuum}) 
of the linear space of vectors obtained by applying arbitrary polynomials in creation operators, 
$a^{*}(\cdot, X),\, X\in \mathcal{X},$ smeared out with test functions in the time variable $t\in \mathbb{R}$. 
The Hamiltonian, $P_0$, of the radiation field has the property that
\begin{equation}\label{field Hamiltonian}
e^{it P_0}\, a^{\#}(s, X)\, e^{-it P_0} = a^{\#}\big(t+s, \phi_{t}(X)\big), \quad \forall\,\, t, s \text{ in } \mathbb{R},\,\, 
\forall \,\,X \in \mathfrak{X}\,,
\end{equation}
where $\mathcal{X}\ni X\mapsto \phi_{t}(X)\in \mathcal{X}, \,t \in \mathbb{R},$ is some deterministic 
dynamics defined on $\mathcal{X}$ that preserves the quadratic form $C$; the choice of $\phi_t$ is 
irrelevant in the following discussion. The spectrum of $P_0$ covers the entire real line and 
is absolutely continuous. As a consequence, there is always some quantum noise caused by the 
radiation field in every state, including the ``vacuum state'' (as if the temperature of radiation were 
positive). 

By $\big|\gamma\big>$ we denote a state of $\geq 1$ photons; any such state is orthogonal to the vacuum, 
i.e., $\big< \gamma \big| \emptyset\big> =0$.
General states of the radiation field are density matrices on $\mathfrak{F}$.

\textbf{Remark}: Simpler models of the radiation field with discrete time have been 
considered in \cite{FP}. The purpose of introducing the model presented above is just to make it
clear that, in the limit where the velocity of light tends to $\infty$, we can actually accommodate models 
with continuous time.

States of the photomultiplier won't appear explicitly in our discussion. The only important feature is
that the state of the ``dormant'' photo-multiplier is \textbf{orthogonal} to all its states right after being hit by 
some photons, i.e., states when it fires. A photo-multiplier is a system with infinitely many degrees of 
freedom with the property that states occupied by the photo-multiplier right after being hit by some ``photons,'' 
when evaluated on ``quasi-local observables,'' relax back to the state of the ``dormant'' photo-multiplier 
within a short relaxation time. Simple models of such systems have been studied; 
(see, e.g., \cite{FGS, DeR-K, FP}).\\

Disregarding the photo-multiplier, the Hamiltonian, $H_S$, of the system to be studied is given by
\begin{equation}\label{total Ham}
H_S= H_A \otimes \mathbf{1} + \mathbf{1}_2 \otimes P_0 + H_I\,,
\end{equation}
where $H_I$ is an \textbf{interaction Hamiltonian} describing emission and absorption of ``photons'' by the atom. For example,
$$H_I= g\,\big[ \sigma_{-}\otimes a^{*}(0, X_0) + \sigma_{+} \otimes a(0, X_0)\big]\,,$$
where $g$ is a real coupling constant assumed to be small, $\alpha:= g^{2}\ll 1$, 
$\sigma_{-}=\begin{pmatrix} 0&0\\1&0 \end{pmatrix}$ is the usual lowering operator on 
$\mathfrak{h}_A$, $\sigma_{+} =\begin{pmatrix} 0&1\\0&0 \end{pmatrix}$ is the 
raising operator, and $X_0\in \mathcal{X}$ is the ``position of the atom.'' The details of how 
$H_I$ is chosen do not matter in what follows next. 

\subsection{Effective time evolution of an atom coupled to the radiation field}
In this Subsection we describe the main results of our analysis; (details will be reported
in a separate paper, but see also \cite{FP}).

In both kinds of experiments described at the beginning of Subsection 5.1 (i.e., when photons emitted by
the atom are not detected, or when they are detected, respectively) the rules of the $ETH$ - Approach described
in Section 3 enable one to eliminate the degrees of freedom of the electromagnetic field and of the photo-multiplier
and come up with an effective time evolution of the atom. When we take an ensemble average over many such atoms,
all of them identically prepared, we obtain a linear, deterministic evolution equation for (what we have called) the
ensemble state, $\rho$, of the two-level atom (ignoring its orbital motion). It is given by the following Lindblad
equation.
\begin{equation}\label{LEq}
\dot{\rho}(\vec{n}_t) = -i \big[H_A, \rho(\vec{n}_t)\big] + \alpha \,\sigma_{-} \cdot \rho(\vec{n}_t)\cdot \sigma_{+} - 
\frac{\alpha}{2} \big\{\rho(\vec{n}_t), \sigma_{+}\sigma_{-}\big\}\,,
\end{equation}
where $\dot{\rho}$ denotes the derivative of $\rho$ with respect to time $t$, and $\alpha = \mathcal{O}(g^{2})$, with
$g$ some coupling constant. This equation actually implies a linear evolution equation for the vector $\vec{n}_t$
with the property that the solution $\vec{n}_t$ stays inside the closed unit ball 
$\big\{ \vec{n} \in \mathbb{R}^{3}\,\big| \,\big|\vec{n}\big| \leq 1\big\}$, provided the initial 
condition, $\vec{n}_0$, belongs to it.

The rules of the $ETH$ - Approach applied to this particular system imply that the \textbf{stochastic time 
evolution} of the state of an \textbf{individual} atom is given by an \textbf{``unraveling''} of the Lindblad evolution 
given by Eq.~\eqref{LEq}. The process of ``unraveling'' Eq.~\eqref{LEq} is analogous to passing from the 
diffusion equation \eqref{diffusion} of Section 2 for the density of particles suspended in a fluid to the 
Brownian motion of a single such particle, as described by the Wiener process. 

There are two distinct ``unravelings'' of Eq.~\eqref{diffusion} corresponding to the two different 
experiments alluded to above. We cannot go into details about how to derive these unravelings from the
rules of the $ETH$ - Approach, as stated in Section 3; but we describe the outcome of our analysis. 
(Complete arguments will be presented in a separate paper, see also \cite{FP}.)

\begin{enumerate}
\item{\textbf{Photo-multiplier turned off}.

The initial state of the radiation field is chosen to be the vacuum $\big|\emptyset\big>$, which 
(in the limit where $c\rightarrow \infty$) does \textbf{not} entangle the states of ``photons'' at 
different times. The effective time evolution of the state of an individual atom is then ``Markovian'' and 
can be determined explicitly. In notations inspired by 
those used in Sect.~2, with $\overline{\omega}_{t+\tau} \mapsto \overline{\vec{n}}(t+dt)$, 
we find that
\begin{align}
\overline{\vec{n}}(t+dt)= \,\vec{n}(t) + d\vec{n}(t)\,, \quad \text{with }\quad
d\vec{n}(t) =\,\vec{\omega}\times \vec{n}(t) \,dt + dK\big[\vec{n}(t)\big]\,.
\end{align}
Here $\vec{\omega}:= (0,0,\Omega)$, and $dK$ is a linear ``dissipative'' map proportional in size to $dt$ 
(and known explicitly) which has the effect that the length of $\big|\vec{n}(t)\big|$ shrinks,
$$|\overline{\vec{n}}(t+dt)| = 1-\mathcal{O}(\alpha)\,dt<1, \quad \text{except if }\,\vec{n}(t)=-\vec{e}_3.$$
Applying \textbf{Axiom CP} of Sect.~3, the evolution of the state of an atom is found to be given by 
a \textbf{Poisson-like jump process} on the Bloch sphere:
\begin{align}\label{jumps}
\begin{split}
{+}) \quad \vec{n}(t) \mapsto &\vec{n}(t+dt):=  \frac{\overline{\vec{n}}(t+dt)}{|\overline{\vec{n}}(t+dt)|} \quad 
\text{with probability } 1-\mathcal{O}(\alpha)\,dt\,, \\
{-}) \quad\vec{n}(t) \mapsto &\vec{n}(t+dt):= \mathbf{-}\,\frac{\overline{\vec{n}}(t+dt)}{|\overline{\vec{n}}(t+dt)|} 
\quad \text{with probability } \mathcal{O}(\alpha)\,dt\,.
\end{split}
\end{align}
The rate of this jump process is proportional to $\alpha (1+ \vec{n}_t\cdot \vec{e}_3)$, i.e., the number of jumps from
$\vec{n}(t)$ to its antipode during the atom's trip from source to detector is proportional to 
$\alpha \,T$. The times when jumps occur are \textbf{random variables} whose law can be determined quite explicitly. 
(In verifying these claims, it may be helpful to first imagine that time is discrete, as in Sects.~2 and 3, with $dt=\tau>0$, 
and to then let $dt$ approach 0 at the end of the calculations.)

When entering the atom detector the state of the atom is given by
\begin{equation}\label{out state}
\rho(\vec{n}_{out}), \quad\text{with } \,\, \vec{n}_{out}\approx 
(\text{sin}\theta_{out}\, \text{cos}\varphi_{out}, \text{sin}\theta_{out}\, \text{sin}\varphi_{out}, \text{cos}\theta_{out})\,,
\end{equation}
where $|\theta_{out} - \theta_0| = \mathcal{O}(\alpha\,T)$ (no jump, mod. 2), $|\theta_{out} -\pi+ \theta_0| = 
\mathcal{O}(\alpha\,T)$ 
(1 jump, mod. 2), assuming that $\alpha\,T \ll 1$.}
\item{\textbf{Photo-multiplier turned on}.

We begin with the observation that, in our model, the dynamics of the photo-multiplier really only enters 
in so far as the state of the ``dormant'' photo-multiplier,  i.e, the state of the photo-multiplier at the moment when 
an atom leaves the atom source, is orthogonal to its state right after being hit by a ``photon'' emitted 
by an atom on its journey from source to detector. For this reason, states of the photo-multiplier 
do not appear explicitly in our formulae and are therefore not indicated in the following.
 
 According to the $ETH$ - Approach, the time evolution of the initial state 
 $$\Psi_{in}:=\rho(\vec{n}_0)\otimes \big|\emptyset\big> \big< \emptyset\big|$$ of the atom and the ``radiation field''
 to the final state when the atom enters the atom detector and a ``photon'' may or may not have been emitted 
 by it is given by 
\begin{equation}\label{full evol} 
\Psi_{in} \mapsto \Psi_{out}^{(0)}= \rho(\vec{n}_{out})\otimes \big|\emptyset\big>\big< \emptyset \big|, \quad 
\text{with probability }\quad 1-\mathcal{O}(\alpha\,T),
\end{equation}
when no photon has been emitted by the atom, where 
$\vec{n}_{out} = (\text{sin}\theta_{out}\, \text{cos}\varphi_{out}, \text{sin}\theta_{out}\, \text{sin}\varphi_{out}, \text{cos}\theta_{out})$,
with $|\theta_{out} - \theta_0| = \mathcal{O}(\alpha\,T)$; and
\begin{equation}\label{outstate}
\Psi_{in}\mapsto \Psi_{out}^{(1)}= \rho(-\vec{e}_3) \otimes \big| \gamma \big>\big< \gamma \big|, \quad 
\text{with probability }\quad \mathcal{O}(\alpha\,T),
\end{equation}
when a photon has been emitted by the atom and recorded by the photon multiplier, where 
$\big| \gamma \big>$ is orthogonal to $\big| \emptyset \big>,$ and 
$\rho(-\vec{e}_3)= |\downarrow\rangle \langle \downarrow|$.}
\end{enumerate}

If $\alpha\,T$ is not very small the difference between the atomic out-states in the \textbf{absence} 
of the photo-multiplier and in its \textbf{presence}, respectively, can be detected by measurements 
of suitable atomic ``observables'' in the atom detector. These measurements, too, can be described 
within the $ETH$- Approach (see \cite{Fr, FP}).

We think that, when compared to earlier treatments of fluorescence (see, e.g., \cite{Pomeau}), 
the analysis sketched here represents progress.

\section{A brief sketch of relativistic quantum theory}
To begin our sketch of relativistic quantum theory, we should recall that relativistic theories, whether 
classical or quantum, can usually \textbf{not} be used by any observers to fully predict the future. 
To provide an example confirming this claim one may consider a space-time with an event horizon. For 
fundamental reasons, observers in such a space-time do not have access to complete information about 
the initial conditions of the Universe. Hence they do \textbf{not} and \textbf{cannot} know of events 
taking place outside their event horizon. However such events, unpredictable as they are for them, 
can and usually will affect their future. 

One may thus characterize the situation of an arbitrary observer by saying that 
\begin{itemize}
\item{his/her past is a history of events or facts that have materialized inside his/her past light cone 
(but are usually only incompletely recorded); while}
\item{his/her future consists of potentialities localized inside his/her future light cone whose 
actualization is not predictable with certainty.}
\end{itemize}
As we have seen in previous sections and will see in the following, this \textbf{Aristotelian dichotomy} 
is realized perfectly in the quantum theory of individual isolated systems.

\subsection{The past-future dichotomy in relativistic quantum theory}
This section follows \cite{Fr}. It is somewhat technical, and readers who do not have much taste for 
slightly subtle mathematical concepts may want to skip it. (Details of the material reviewed here
will appear elsewhere.)

We temporarily suppose that space-time is given by Minkowski space, $\mathbb{M}^{4}$, 
with gravity turned off. This does, of course, not do justice to the full story concerning relativistic 
local quantum theory. But a complete story is not known in precise detail, yet, and what is known of it 
is rather too lengthy to be reviewed here.

For $P\in \mathbb{M}^{4}$, $V^{+}_{P}$ denotes the forward (future) light cone 
and $V^{-}_{P}$ the backward (past) light cone with apex in $P$.
A ``diamond,'' $D$, is defined by 
$$D\equiv D_{P, P'}: = V^{+}_{P} \cap V^{-}_{P'}, \,\text{ with }\,\,P' \in V^{+}_{P}\,.$$
With every diamond $D\subset \mathbb{M}^{4}$ we associate an algebra, $\mathfrak{E}(D)$, (a von 
Neumann algebra, to be precise) generated by \textbf{``potentialities''} that are represented by partitions 
of unity by orthogonal projection operators acting on a separable Hilbert space $\mathcal{H}$ and localized in $D$. 
If two diamonds, $D$ and $D'$, are space-like separated then the property of \textbf{locality} 
(or Einstein causality) says that
$$[A, B] = 0, \qquad \forall \,\,\, A \in \mathfrak{E}(D)\, \text{ and }\,\,B\in \mathfrak{E}(D')\,.$$
The algebras $\mathfrak{E}(D)$ are called \textbf{local algebras}. Typically operators in 
$\mathfrak{E}(D)$ consist of bounded functions of local field operators smeared out with 
test functions whose support is contained in $D$. The net 
$$\big\{\mathfrak{E}(D)\,\big|\, D\subset \mathbb{M}^{4}\big\}$$
of local algebras is assumed to have all the usual properties -- locality, covariance, etc. -- assumed in algebraic 
quantum field theoy (AQFT), see \cite{Haag}, and, in addition, \textbf{Landau's property} (see\cite{Landau}), 
which, under suitable extra hypotheses, can be derived from the basic assumptions of AQFT and says that
\begin{equation}\label{Landauprop}
\text{if }\,\,\,D_1 \cap D_2 = \emptyset \quad\text{then}\quad \mathfrak{E}(D_1) \cap \mathfrak{E}(D_2) = \emptyset\,.
\end{equation}
Given a forward light cone $V^{+}\subset \mathbb{M}^{4}$, we define
$$\mathfrak{E}(V^{+}):= \bigvee_{D\subset V^{+}} \mathfrak{E}(D)\,.$$
We note that $\mathfrak{E}(V^{+})$ is \textbf{not} defined to be norm-closed. Landau's property then implies that
\begin{equation}\label{empty-int}
\text{if }\,\,\,D\cap V^{+} =\emptyset \quad\text{then}\quad \mathfrak{E}(D)\cap \mathfrak{E}(V^{+}) = \emptyset\,.
\end{equation}
We expect that, for theories with massless modes, \eqref{empty-int} holds even if $\mathfrak{E}(V^{+})$ were replaced by 
its weak closure; (a property that -- if true -- might be useful in a general analysis of relativistic quantum theory.) 

Potential events, $\mathfrak{P}$, are defined to be partitions of unity by disjoint orthogonal projections,
$\pi_{\xi} \in \mathfrak{E}(D)$, for some diamond $D$,
$$\mathfrak{P}=\big\{\pi_{\xi}\,\big|\, \pi_{\xi} \in \mathfrak{E}(D), \,\xi \in \mathfrak{X}_{\mathfrak{P}}\big\}, \quad\text{with}\quad \pi_{\xi}=\pi_{\xi}^{*},\,\,
\pi_{\xi}\cdot \pi_{\eta}= \delta_{\xi \eta} \pi_{\xi}, \,\forall \,\, \xi, \eta \,\,\text{ in }\,\,\mathfrak{X}_{\mathfrak{P}}\,,$$
where $\mathfrak{X}_{\mathfrak{P}} \equiv \text{spec}(\mathfrak{P})$ is a countable
set of indices called the \textbf{spectrum} of the potential event $\mathfrak{P}$.
\vspace{0.1cm}
We define the \textbf{``support,''} $D_{\mathfrak{P}}$, of a potential event $\mathfrak{P}$ 
to be the smallest diamond with the property that 
$\mathfrak{P} \subset \mathfrak{E}(D_{\mathfrak{P}})$.
Given a potential event $\mathfrak{P}$ with support $D=D_{\mathfrak{P}}$, we define
$V^{+}_{\mathfrak{P}}$ to be the smallest forward light cone containing $D_{\mathfrak{P}}$, and we define
$V^{-}_{\mathfrak{P}}$ to be the backward light cone with the same apex as $V^{+}_{\mathfrak{P}}$, i.e., 
the backward light cone in the past of $D_{\mathfrak{P}}$.
To shorten our notations we set
\begin{equation}\label{notations}
\mathfrak{E}_{\mathfrak{P}}:= \mathfrak{E}(D_{\mathfrak{P}}), \qquad
\mathfrak{E}^{+}_{\mathfrak{P}}:= \mathfrak{E}(V^{+}_{\mathfrak{P}})\,.
\end{equation}

Next, we consider \textbf{causal relationships} between different potential events, and we introduce a
convenient notion of state.\\

\noindent
\textbf{Definition}:
\begin{enumerate}
\item[(i)]{A potential event $\mathfrak{P}$ is in the \textbf{past} of another potential event $\mathfrak{P}'$, 
written as $\mathfrak{P}\prec \mathfrak{P}'$, iff
\begin{align}\label{past}
\mathfrak{E}_{\mathfrak{P}'} \subset \mathfrak{E}^{+}_{\mathfrak{P}} 
\end{align}
If $\mathfrak{P} \prec \mathfrak{P}'$ we say that $\mathfrak{P}'$ is in the \textbf{future} of $\mathfrak{P}$,
written as $\mathfrak{P}' \succ \mathfrak{P}$.

We note that if $D_{\mathfrak{P}'} \cap V^{+}_{\mathfrak{P}} = \emptyset$ then 
$\mathfrak{E}_{\mathfrak{P}'} \cap \mathfrak{E}^{+}_{\mathfrak{P}} = \emptyset$; hence 
$\mathfrak{P} \not\prec \mathfrak{P}'$ (i.e., $\mathfrak{P}$ is not in the past of $\mathfrak{P}'$),
as follows from Landau's property.}
\item[(ii)]{ If \,\,$\mathfrak{P} \not\prec \mathfrak{P}'$ and $\mathfrak{P}' \not\prec \mathfrak{P}$\,
and if all operators in the algebra $\mathfrak{E}_{\mathfrak{P}}$ commute with all operators in 
$\mathfrak{E}_{\mathfrak{P}'}$\,\, we say that $\mathfrak{P}$ and $\mathfrak{P}'$ 
are \textbf{space-like separated}, written as $\mathfrak{P} \bigtimes \mathfrak{P}'$.}
\item[(iii)]{Let $\Sigma$ be the intersection of a space-like surface with the interior of a backward light cone 
$V^{-}$; (i.e., $\Sigma$ is the piece of a space-like surface contained in the interior of $V^{-}$.) 
A \textbf{state}, $\omega_{\Sigma}$, on the potential events in the future of $\Sigma$ is a normal 
state on the algebra
\begin{equation}\label{Sigma}
\mathfrak{E}^{+}_{\Sigma}:= \bigvee_{P\in \Sigma} \mathfrak{E}(V^{+}_P)\,.
\end{equation}
}
\end{enumerate}
We define $V^{-}_{\Sigma}$ to be the set of space-time points $P \in V^{-}$ contained in the future of
$\Sigma$.
We specify, in the form of two postulates, \textbf{necessary conditions} for a family of potential events, 
$$\mathfrak{F}_{\omega_{\Sigma}}:=\big\{ \mathfrak{P} \subset \mathfrak{E}^{+}_{\Sigma}\,\big|\, D_{\mathfrak{P}} 
\subset V^{-}_{\Sigma} \big\}$$ 
localized in $V^{-}_{\Sigma}$ to \textbf{actualize}, given a state $\omega_{\Sigma}$ on the algebra $\mathfrak{E}^{+}_{\Sigma}$. 

\subsection{Basic postulates of relativistic quantum theory}

\hspace{0.6cm}$\bullet$ \textbf{Postulate 1}: For any two distinct potential events $\mathfrak{P}$ and $\mathfrak{P}'$ in 
$\mathfrak{F}_{\omega_{\Sigma}}$, one of the following must hold:
\begin{equation}\label{causalrelation}
\mathfrak{P} \prec \mathfrak{P}', \quad \text{or} \quad \mathfrak{P}' \prec \mathfrak{P}\,, \quad \text{or} 
\quad \mathfrak{P} \bigtimes \mathfrak{P}'\,.\vspace{0.2cm}
\end{equation}

If $\mathfrak{P} \in \mathfrak{F}_{\omega_{\Sigma}}$ is a potential event that actualizes 
then a variant of the state-reduction postulate, \textbf{Axiom CP}, formulated in Subsection 3.3 says 
that there is an orthogonal projection $\pi \in \mathfrak{P}$ that \textbf{materializes}, in the sense 
that the state on the algebra $\mathfrak{E}^{+}_{\mathfrak{P}}$ is an \textbf{eigenstate} of $\pi$ (i.e., lies in the
range of $\pi$). Such a projection $\pi$ is called an \textbf{actuality} or \textbf{fact}. 
Given a potential event $\mathfrak{P}\in \mathfrak{F}_{\omega_{\Sigma}}$, 
the \textbf{past history}, $\mathfrak{H}_{\mathfrak{P}}$, of $\mathfrak{P}$ consists of all the 
actualities (or facts), $\pi' \in \mathfrak{P}'$, that have materialized, where the potenial event 
$\mathfrak{P}' \in \mathfrak{F}_{\omega_{\Sigma}}$ is such that $D_{\mathfrak{P}'}\subset V^{-}_{\mathfrak{P}}$, 
i.e., $\mathfrak{P}' \prec \mathfrak{P}$. 
With a history $\mathfrak{H}_{\mathfrak{P}}$ in the past of $\mathfrak{P}$ we associate a 
\textbf{``history operator''} defined by
\begin{equation}\label{historyops}
H_{\mathfrak{P}}:= \underset{\pi' \in \mathfrak{H}_{\mathfrak{P}}}{{\overset{\rightarrow}{\prod}}} \pi'\,,
\end{equation}
where $\overset{\rightarrow}{\prod}$ is an ordered product over actualities 
$\pi' \in \mathfrak{P}' \in \mathfrak{F}_{\omega_{\Sigma}}$ with the property 
that if $\mathfrak{P}' \prec \mathfrak{P}'' \in \mathfrak{F}_{\omega_{\Sigma}}$ then $\pi'$ stands to the left of $\pi''$; (if 
$\mathfrak{P}' \bigtimes \mathfrak{P}''$ then the order of $\pi'$ and $\pi''$ does not matter, 
because, in a \textbf{local} relativistic quantum theory, these two operators commute!).

We define a state, denoted by $\omega_{\mathfrak{P}}$, on the algebra $\mathfrak{E}^{+}_{\mathfrak{P}}$ 
by setting
\begin{equation}\label{state}
\omega_{\mathfrak{P}}(X):= \mathcal{Z}^{-1}_{\mathfrak{P}}\,\omega_{\Sigma}\big(H_{\mathfrak{P}}\cdot X\cdot
H^{*}_{\mathfrak{P}}\big), \qquad \forall \,\,\,X\in \mathfrak{E}^{+}_{\mathfrak{P}}\,,
\end{equation}
where $\mathcal{Z}_{\mathfrak{P}}= \omega_{\Sigma}(H_{\mathfrak{P}}\cdot H^{*}_{\mathfrak{P}})$ is a 
normalization factor chosen such that $\omega_{\mathfrak{P}}(\mathbf{1}) =1$.

$\bullet$ \textbf{Postulate 2}: Given the state $\omega_{\mathfrak{P}}$, the potential event $\mathfrak{P}$ \textbf{actualizes} 
iff it generates the \textbf{center}, $\mathcal{Z}(\omega_{\mathfrak{P}})$, of the centralizer, 
$\mathcal{C}(\omega_{\mathfrak{P}})$, of the state $\omega_{\mathfrak{P}}$ and 
$\mathcal{Z}(\omega_{\mathfrak{P}})$ is non-trivial; (see Subsection 3.2 for definitions).

The state reduction postulate (see Axiom CP, of Subsection 3.3) is now formulated as follows.

$\bullet$ \textbf{Axiom CP:}
If, according to Postulate 2, the potential event $\mathfrak{P}$ actualizes then some \textbf{actuality}
or fact, $\pi \in \mathfrak{P}$, materializes.\\

The feasibility of Postulate 2 -- namely the possible existence of a non-trivial algebra 
$\mathcal{Z}(\omega_{\mathfrak{P}})$ -- can be understood as a consequence of (a generlization of) 
the \textbf{Principle of Diminishing Potentialities} (see item (IV) of Subsection 3.1), which 
can be derived from a basic result proven in \cite{Buchholz}: 
If (and only if) the theory describes massless particles, such as photons (and/or gravitons), then, for 
a space-time point $P'$ in the future of a point $P$, the relative commutant 
\mbox{$\mathfrak{E}(V^{+}_{P'})' \cap \mathfrak{E}(V^{+}_{P})$} (see Subsection 4.1, 
right above \eqref{rel commutants}) is an \textbf{infinite-dimensional} algebra. This is what 
has been called \textbf{Huygens' Principle} in Section 4, following \cite{Buchholz}; (see also \cite{BF}). 
Apparently, Huygens' Principle implies (a suitably generalized form of) the Principle of Diminishing Potentialities!\\

Next, we discuss what we call the relativistic Born Rule. Given the data 
$\big\{V^{-}, \Sigma, V^{-}_{\Sigma}, \omega_{\Sigma}\big\}$ and $\mathfrak{F}_{\omega_{\Sigma}}$, 
as specified above, we define $\mathfrak{H}\big(\mathfrak{F}_{\omega_{\Sigma}}\big)$ 
to be the history of actualities or facts $\pi \in \mathfrak{P}\in \mathfrak{F}_{\omega_{\Sigma}}$ 
localized in $V^{-}_{\Sigma}$ that have materialized. With a history 
$\mathfrak{H}\big(\mathfrak{F}_{\omega_{\Sigma}}\big)$ we associate the history operator
\begin{equation}\label{hist-op}
H_{\mathfrak{F}_{\omega_{\Sigma}}}:= \underset{\pi\in \mathfrak{H}(\mathfrak{F}_{\omega_{\Sigma}})}
{\overset{\rightarrow}{\prod}} \pi
\end{equation}
\textbf{Relativistic Born Rule:}
The probability, $\text{prob}\big[\mathfrak{H}\big(\mathfrak{F}_{\omega_{\Sigma}}\big)\big]$, 
of a history $\mathfrak{H}\big(\mathfrak{F}_{\omega_{\Sigma}}\big)$, given that its future is unknown,
is determined by the formula
\begin{equation}\label{prob-history}
\text{prob}\big[\mathfrak{H}\big(\mathfrak{F}_{\omega_{\Sigma}}\big)\big]:=
\omega_{\Sigma}\big(H_{\mathfrak{F}_{\omega_{\Sigma}}}\cdot H^{*}_{\mathfrak{F}_{\omega_{\Sigma}}}\big)\,.
\end{equation}
Clearly $\text{prob}\big[\mathfrak{H}\big(\mathfrak{F}_{\omega_{\Sigma}}\big)\big]$ is non-negative, and
$$\underset{\pi\in \mathfrak{H}(\mathfrak{F}_{\omega_{\Sigma}})}{\sum} 
\text{prob}\big[\mathfrak{H}\big(\mathfrak{F}_{\omega_{\Sigma}}\big)\big] = 1\,.$$
\textbf{Remarks}:
\begin{enumerate}
\item[(i)] {Note that the conditional probability of a history $\mathfrak{H}\big(\mathfrak{F}_{\omega_{\Sigma}}\big)$ 
predicted by the theory changes as one moves into its future by embedding it into a longer history. 
However, if the probabilities of longer histories are summed over all possible actualities or facts that 
may materialize in the course of \textbf{future} evolution, i.e., in the future of the given history 
$\mathfrak{H}\big(\mathfrak{F}_{\omega_{\Sigma}}\big)$ or space-like to it, then one recovers 
expression \eqref{prob-history}, as is easily verified.}
\item[(ii)] {For an arbitrary actuality $\pi_0 \in \mathfrak{P}$, we have that
$$\underset{\pi \in \mathfrak{H}_{\mathfrak{P}}}{\sum} \omega_{\Sigma}\big(H_{\mathfrak{P}}\,(\pi_0\,X\, \pi_0)\,
H^{*}_{\mathfrak{P}}\big) =\omega_{\Sigma}\big(\pi_0\,X\, \pi_0\big)\,, \qquad \forall \,\,\, X\in 
\mathfrak{E}^{+}_{\mathfrak{P}}\,.$$ This implies that \textbf{``ensemble states''} do not evolve.}
\item[(iii)] {The \textbf{ontology} of relativistic quantum theory lies in histories, $\mathfrak{H}$, of 
actualities satisfying Postulates 1 and 2 and Axiom CP.}
\item[(iv)] {Postulates 1 and 2 impose very strong, seemingly \textbf{non-local} constraints on the
actualization of potential events. The significance and consequences of these postulates will have  
to be analyzed more deeply than has been done, so far.}
\item[(v)]{One appears to be capable to eliminate any reference to a specific model of space-time and to
reformulate everything in this Section in terms of relations between algebras $\mathfrak{E}_{\mathfrak{P}}$ 
and $\mathfrak{E}^{+}_{\mathfrak{P}}$, for different potential events $\mathfrak{P}$. In particular, 
Postulates 1 and 2 and Axiom CP can all be formulated without referring to a specific model of space-time!
In other words, one appears to be able to eliminate geometrical concepts and replace them by purely 
algebraic ones and by relations between different algebras that express causal relationships between 
potential events and actualities. The picture then emerges that space-time is ``woven'' through 
the emergence of actualities or facts, as determined by Postulates 1 and 2 and Axiom CP. 
While a sketch of this project has been worked out we refrain from describing it here.}

\item[(vi)]{Postulates 1 and 2 and Axiom CP \textbf{cannot} be expected to recursively determine the events 
actualizing inside any past light cone; (in contrats to what we have seen in the non-relativistic framework 
described in Section 3). In view of what has been said above and in items (iv) and (v), we expect the resulting 
ambiguities to express effects of \textbf{gravity}. Of course, details remain to be understood more precisely.}
\end{enumerate}
A more detailed account of these ideas will appear elsewhere (but see also \cite{Fr}).

\section{Conclusions and ackonwledgments}
Here are some general conclusions derived from the insights into QM reviewed in this paper.
\begin{enumerate}
\item{The \textit{ETH} - Approach to Quantum Mechanics represents a \textbf{completion of QM} that provides
a precise description of the \textbf{stochastic time evolution} of states of \textbf{individua}l 
systems unraveling Schr\"odinger - von Neumann evolution, and of \textbf{events} and their recordings; 
see also \cite{BFS, Froh, Fr}. It has resemblences, albeit rather vague ones, with Everett's ``Many Worlds'' 
formalism \cite{Everett} and spontaneous collapse models \`a la ``GRW'' \cite{GRW}. But it supersedes 
these ad-hoc formalisms by a more natural and intrinsic framework. Of course, it will have to stand the 
test of experiments, as every new law of Nature one proposes has to.}
\item{To quote Wolfgang \textit{Pauli}: \textit{If speculative ideas cannot be tested, they're not science; 
they don't even rise to the level of being wrong.}
We thus should ask whether the \textbf{Principle of Diminishing Potentialities} (PDP), which is a 
corner stone of the $ETH$ - Approach to QM, is more than a speculative idea and whether 
it can be tested. It is clear that this principle can only be valid for quantum theories of systems 
with infinitely many degrees of freedom. (For example, it does not hold in a model of non-relativistic matter
not interacting with the quantized electromagnetic field.) It has the status of a \textbf{theorem} in local 
relativistic quantum theories with \textbf{massless} particles on \textbf{even-dimensional} space-times; 
e.g., in 4D quantum electrodynamics (QED); see \cite{Buchholz} and \cite{BF}. It also holds in simple 
models of QED regularized at high energies by discretizing time, which we sketched in Section 4 (see \cite{FP} 
for further details); and it holds in models emerging in the limit of the velocity of light tending to $\infty$,
as described in Section 5 and in \cite{FP}. However, in this limit, the Hamiltonian of the total system 
is not bounded from below; i.e., the usual spectrum condition ($\nexists$ negative-energy states) is violated. 

We thus have strong reasons to expect that a completion of QM without negative-energy states that
solves the so-called \textbf{``measurement problem''} will only succeed in the guise of
\textbf{local relativistic quantum theory} on even-dimensional space-times with massless 
bosons, photons and gravitons, so that, as a consequence of ``Huygens' Principle,'' the Principle of
Diminishing Potentialities (PDP) holds. Besides Huygens' Principle there may, however, be other 
physical mechanisms implying PDP. For example, certain theories with extra dimensions or theories
predicting an abundant generation of black holes may exhibit such mechanisms. 

Recasting the material presented in Section 6 in a purely algebraic manner not referring to a
specific back-ground space-time might be a promising starting point towards understanding
how \textbf{gravity} enters our formalism and how the actualization of potential events ``weaves''
space-time.

We regard these observations as important novel insights into the consistency of a framework 
underlying a completion of quantum theory and eliminating the puzzles of text-book QM.}

\item{We have applied the non-relativistic $ETH$ - Approach, as developed in Section 3, to
models of \textbf{measurements of physical quantities} characteristic of a subsystem of an isolated 
system; see \cite{FP}. The success has been striking. 

Furthermore, in \cite{FSchub}, models describing the preparation of specific initial states of isolated systems
have been discussed in the spirit of the $ETH$ - Approach.}

\item{The $ETH $- Approach to QM, in particular PDP, introduces a fundamental distinction 
between past and future into the theory: The past consists of \textbf{facts}, namely histories of 
\textbf{``actualities''}, while the future consists of \textbf{``potentialities''} (much in the sense in which
\textit{Aristotle} originally conceived these concepts). This feature and the \textbf{``Arrow of Time''} intrinsic in
QM have been highlighted in \cite{Arrow}.}
\end{enumerate}

To conclude, we wish to express our hope that the ideas described in this paper will be taken
seriously and widely discussed among people interested in the foundations of QM.\\

\textbf{Acknowledgements:} One of us (J.~F.) thanks his collaborators in earlier work related to the
foundations of QM, in particular Baptiste \textit{Schubnel,} for the pleasure of cooperation. He is grateful
to Carlo \textit{Albert,} Philippe \textit{Blanchard,} Detlev \textit{Buchholz,} David \textit{Ellwood,} Shelly 
\textit{Goldstein} and Erhard \textit{Seiler} for their very encouraging interest in our efforts and many 
helpful discussions. J.~F.~thanks Heinz \textit{Siedentop} for having invited him to teach a short course 
on quantum theory at LMU-Munich, back in November of 2019. This was a great opportunity to experiment
with some of the material presented in this paper; he thanks Heinz for his generous hospitality.

\noindent
1 Institut f\"ur theoretische Physik, ETH-Zurich, 8093 Zurich, Switzerland. Email: juerg@phys.ethz.ch\\
2 Binghamton University, Department of Mathematics and Statistics. Email: gangzhou@binghamton.edu\\
3 Dipartimento di Matematica, Università di Roma ``TorVergata,'' Rome, Italy. Email: pizzo@mat.uniroma2.it
\end{document}